\documentclass[fleqn,usenatbib]{rasti}

\usepackage{newtxtext,newtxmath}

\usepackage[T1]{fontenc}

\DeclareRobustCommand{\VAN}[3]{#2}
\let\VANthebibliography\thebibliography
\def\thebibliography{\DeclareRobustCommand{\VAN}[3]{##3}\VANthebibliography}

\usepackage{graphicx}	
\usepackage{amsmath}	
\usepackage{hyperref}
\usepackage{tikz}
\usepackage{xspace}
\usepackage{fdsymbol}
\usetikzlibrary{shapes.geometric, arrows, external}

\usepackage{scalerel}
\usepackage{comment}

\usepackage[table]{xcolor}

\usepackage[utf8]{inputenc}

\usepackage[switch, modulo]{lineno}

\usepackage{euclid}

\title[Matching Rubin alerts to \Euclid]{\Euclid: An automated system to match Rubin transient alerts to \Euclid observations\thanks{This paper is published on behalf of the Euclid Consortium.}}

\author[C.~Duffy et al.]{
\fontsize{10pt}{10pt}\selectfont
C.~Duffy$^{1}$\thanks{E-mail: c.j.duffy@lancaster.ac.uk},
I.~M.~Hook$^{1}$,
C.~M.~Gutierrez$^{2,3}$,
K.~Paterson$^{4}$,
V.~Petrecca$^{5,6}$,
T.~J.~Moriya$^{7,8,9}$,
F.~Poidevin$^{2,3}$
\newauthor\fontsize{10pt}{10pt}\selectfont
R.~Kotak$^{10}$,
B.~Altieri$^{11}$
% ...
\\
\\
\emph{Affiliations can be found at the end of the article.}
}
\date{Accepted 2026 February 17. Received 2026 February 12; in original form 2026 January 12.}

\pubyear{\the\year{}}

\begin{document}
\definecolor{58712c6c-1b6d-5716-9834-102aad6341dc}{RGB}{175, 179, 255}
\definecolor{f3551e38-74df-57e2-b793-83d7fe876c85}{RGB}{0, 0, 0}
\definecolor{0b71a967-1f15-55a5-9bb9-70efa7b4fc58}{RGB}{51, 51, 51}
\definecolor{747aec21-333b-59ee-84e3-ddff893e5ccd}{RGB}{255, 216, 176}
\definecolor{70af333d-4869-5f2a-97ca-9904a9fce6c3}{RGB}{179, 254, 174}
\definecolor{adaee4ed-88c8-5b21-a9e7-31316ebef86f}{RGB}{255, 179, 178}
\definecolor{5856d031-3da1-575c-834e-c77e9e438c62}{RGB}{162, 177, 195}

\tikzstyle{7f27e395-8473-5b8a-b568-31425bce482a} = [trapezium, trapezium left angle=70, trapezium right angle=110, minimum width=60pt, minimum height=1pt, text centered, text width=1.5cm, font=\small, color=0b71a967-1f15-55a5-9bb9-70efa7b4fc58, draw=f3551e38-74df-57e2-b793-83d7fe876c85, line width=1, fill=58712c6c-1b6d-5716-9834-102aad6341dc]
\tikzstyle{69bbb168-da59-5865-902f-94e77902bf95} = [rectangle, minimum width=50pt, minimum height=1pt, text centered, text width=1.5cm, font=\small, color=0b71a967-1f15-55a5-9bb9-70efa7b4fc58, draw=f3551e38-74df-57e2-b793-83d7fe876c85, line width=1, fill=747aec21-333b-59ee-84e3-ddff893e5ccd]
\tikzstyle{b0e005f8-e267-5a3a-a638-881fb9faed1d} = [diamond, minimum width=50pt, minimum height=2cm, text centered, text width=1.5cm, font=\small, color=0b71a967-1f15-55a5-9bb9-70efa7b4fc58, draw=f3551e38-74df-57e2-b793-83d7fe876c85, line width=1, fill=70af333d-4869-5f2a-97ca-9904a9fce6c3]
\tikzstyle{b4eeb85a-e852-5337-aaeb-b1b7ed281176} = [rectangle, minimum width=50pt, minimum height=1cm, text centered, text width=1.5cm, font=\small, color=0b71a967-1f15-55a5-9bb9-70efa7b4fc58, draw=f3551e38-74df-57e2-b793-83d7fe876c85, line width=1, fill=747aec21-333b-59ee-84e3-ddff893e5ccd]
\tikzstyle{574b93ce-5d31-5729-9ed0-9a79e73d4725} = [rectangle, minimum width=50pt, minimum height=1cm, text centered, font=\small, color=0b71a967-1f15-55a5-9bb9-70efa7b4fc58, draw=f3551e38-74df-57e2-b793-83d7fe876c85, line width=1, fill=747aec21-333b-59ee-84e3-ddff893e5ccd]
\tikzstyle{512bdd77-c3aa-5669-a956-85f7a90c6fb4} = [rectangle, rounded corners, minimum width=30pt, minimum height=1cm, text centered, text width=1.5cm, font=\small, color=0b71a967-1f15-55a5-9bb9-70efa7b4fc58, draw=f3551e38-74df-57e2-b793-83d7fe876c85, line width=1, fill=adaee4ed-88c8-5b21-a9e7-31316ebef86f]
\tikzstyle{7be24b85-97d0-5b76-ba9e-d94005dca8f2} = [thick, draw=5856d031-3da1-575c-834e-c77e9e438c62, line width=2, ->, >=stealth]

\label{firstpage}
\pagerange{\pageref{firstpage}--\pageref{lastpage}}
\begin{titlepage}
\maketitle

\begin{abstract}
The Vera C. Rubin observatory is expected to produce 10 million transient alerts per night in \textit{ugrizy} filters, whilst \Euclid is a visible to near-infrared space telescope engaged in a wide field survey. We present a prototype system to automatically match the transient alerts from Rubin to \Euclid observations. The system produces joint light-curves containing both visible and near-infrared photometry, and joint image cutouts. Using Zwicky Transient Facility alerts as a proxy for Rubin, we demonstrate the system in use in cases where \Euclid did and did not detect the transient and highlight the value that can be added in each case. For transients detected by \Euclid these benefits include identifying the supernovae (SNe) in observations taken prior to ground-based detection, thereby better constraining the explosion time, such as SN 2024pvw detected $\sim3\,\rm d$ prior to ground based detections. In cases where \Euclid did not detect the transient, we demonstrate the benefit of adding \Euclid observations to improve host morphology measurements and associations. 
\end{abstract}

\begin{keywords}
data methods -- software -- techniques: photometric -- supernovae: general -- surveys -- supernovae: individual:(SN 2024pvw, AT 2024pcm, SN 2024aebt)
\end{keywords}
\end{titlepage}


\section{Introduction}

In order to fully understand the properties of astrophysical objects it is often necessary to study them across a range of wavelengths. In the specific case of transients, covering both the optical and near-infrared regimes can reveal information on the energy source and radiative processes in supernovae \citep[SNe;][]{2020NatAs...4..188G}, facilitating the discovery of higher redshift transients and dusty transients \citep{2009ApJ...704..306K,2011A&ARv..19...43G,2024ApJ...961..211M,2025ApJ...979..250D}, and revealing distant red galaxies as potential hosts \citep{Q1-SP016}. In the context of future, wide-area sky surveys, there is a recognised synergy between the wavelength ranges and spatial resolution of the \Euclid mission \citep{2024EuclidI} and the Vera C. Rubin Observatory (hereafter Rubin Observatory). \Euclid is a 1.2\,m space based telescope that began science operations in February 2024 and is designed to survey the sky over a 6 year period. \Euclid carries two main science instruments: the visible imaging instrument, VIS, and the near-infrared spectrometer and photometer, NISP. The Rubin Observatory in Chile consists of an 8.4\,m Telescope that will carry out the Legacy Survey of Space and Time \citep[LSST;][]{2019ApJ...873..111I}. The LSST program will take up about 90\% of its observing time, and  will cover $18\,000$ deg\textsuperscript{2} of sky 800 times over 10 years with \textit{ugrizy} filters, with the remaining 10\% dedicated to higher cadence and ``deep drilling'' fields (two of which are specifically planned to overlap with \Euclid). The synergies between these surveys, both for transients and more broadly, are discussed in \citet{2017ApJS..233...21R}. More recently, \citet{2022zndo...5836022G} expands upon this with detailed science cases and recommendations for implementation of joint data products. In the following subsections we describe the \Euclid and LSST surveys and introduce the goals of our work towards some of the more urgent joint data products for transients.

\subsection{\Euclid}\label{sec:Euclid}
\begin{figure*}
    \centering
    \includegraphics[width=\linewidth]{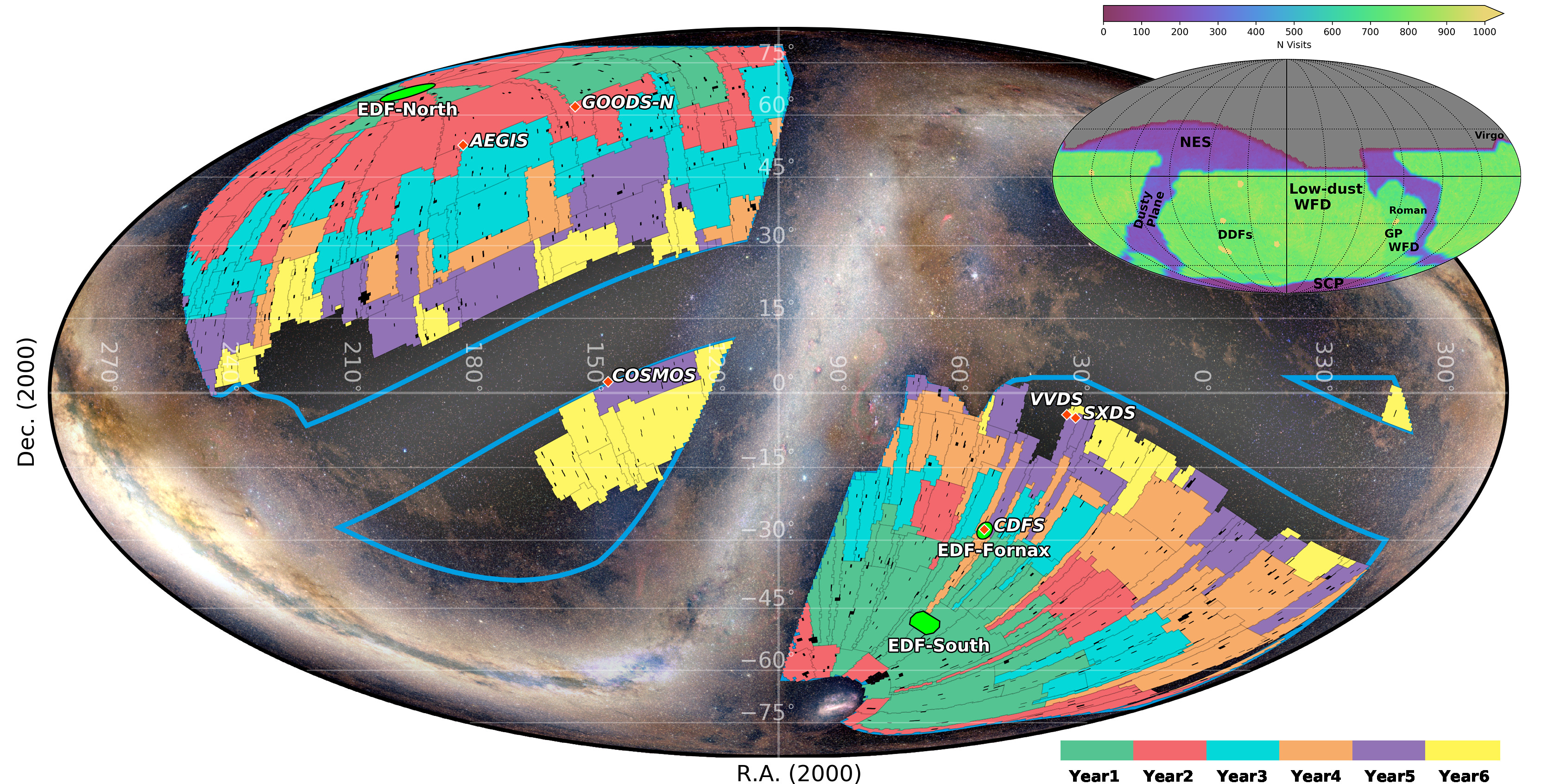}
    \caption{ Image: The Euclid Wide Survey shown in a Mollweide projection. The blue border bounds the entire \Euclid region of interest. Colour coding denotes the year of planned observation within the 6 year survey. Bright green areas denote the Euclid Deep Fields. Red marks denote auxiliary fields. \citep[Figure 25][]{2024EuclidI}. Inset image: The Rubin Observatory’s v5.0.0 10 baseline survey coloured by visits, with special fields such as deep drilling fields (DDF) indicated in text \citep{lsst-survey-plan}.}
    \label{fig:euclid-sky-map}
\end{figure*}

The primary science goals of the \Euclid mission are to probe the nature of dark energy and dark matter using the techniques of weak gravitational lensing and galaxy clustering. In order to do this, \Euclid will survey $14\,000\,\rm deg^{2}$ of the sky in its primary wide field survey (Euclid Wide Survey; EWS, see Fig.~\ref{fig:euclid-sky-map}) using both the VIS and NISP instruments, with each survey field being visited only once over a 6 year period \citep[see][for a full overview of the \Euclid mission]{2024EuclidI}. VIS operates in a single red passband, $I_{\scriptscriptstyle\rm{E}}$, (550--920$\,\rm nm$) with a pixel scale of 0\arcsecf1 pixel\textsuperscript{$-1$} and a limiting AB magnitude of 25 \citep{EuclidSkyVIS}. Imaging from the NISP instrument has three different passbands; $Y_{\scriptscriptstyle\rm{E}}$ (949.6--1212.3$\,\rm nm$), $J_{\scriptscriptstyle\rm{E}}$ (1167.6--1567.0$\,\rm nm$), and $H_{\scriptscriptstyle\rm{E}}$ (1521.5--2021.4$\,\rm nm$) each with a pixel scale of 0\arcsecf3 pixel\textsuperscript{$-1$} and a limiting magnitude of $\sim24.5$ \citep{EuclidSkyNISP}. 

Although the majority of the \Euclid survey will view each part of the sky only once, there is still significant scope to undertake transient science with \Euclid. There are many areas in which \Euclid observations can add scientific value for transients such as identifying or improving host galaxy associations and adding useful early/late time near-infrared (NIR) observations to transients previously identified by ground-based telescopes, which can assist in understanding the progenitors of these systems. \citet{Q1-SP002} demonstrated the value that could be added using single epoch \Euclid observations by manually matching \Euclid observations to transients reported by other surveys to the Transient Name Server (TNS)\footnote{\href{wis-tns.org}{wis-tns.org}}. In that work, \Euclid contributed early time observations of two SN, one of which is one of the earliest NIR observations of a supernova ever made ($\sim15\,\rm d$ prior to peak). Additionally, several transients which had previously been identified as ``orphans'', i.e. without a known host galaxy, were able to be associated with a host galaxy with the \Euclid observations. Adding NIR observations from \Euclid is also useful for observing particularly red or ``dusty'' transients that other surveys cannot detect; and estimating photometric redshift and extinction.

In addition to the wide survey, the \Euclid survey plan includes three deep fields (EDFs; Euclid Deep Fields) which are subject to repeated observations throughout the survey, with approximately 6 visits per year \citep{2022A&A...662A.112E}. Though the cadence of the EDFs is sparse, these are valuable as many observations of e.g., Type Ia and core-collapse supernovae (SNe) are expected \citep{Bailey23}. Furthermore it is expected that a substantial number of high-redshift superluminous and pair-instability SNe will be discovered within the the EDFs \citep{Inserra18,Moriya22,Tanikawa23,briel2024}. The EDFs are shown in Fig\,\ref{fig:euclid-sky-map}, overlaid upon the EWS. Work is ongoing to generate transient detections from \Euclid observations, with a transient detection pipeline currently in the advanced stages of development; this ingests raw \Euclid observations from the EDFs, reduces them and performs image subtraction. Several transients have been identified through this pipeline \citep[e.g. AT 2023adqt,][]{2024TNSTR2306....1C}. It is expected that this work will ultimately result in the development of a dedicated \Euclid transients alert stream.

\subsection{LSST and Lasair}

It is expected that Rubin, through LSST, will be able to detect of order 10 million transient events per night due to its wide field of view and the photometric depth that it can achieve. A single epoch of observation with Rubin will be capable of reaching photometric depths of $r_{\scriptscriptstyle\rm{Rubin}} < 24.5$ with a anticipated cadence of $\sim 2$--4~d in the main survey \citep{2019ApJ...873..111I,2022ApJS..258....1B}.

Transient detections and related information from Rubin are to be released in the form of transients alerts, which are small, rapidly produced data packets. Primarily, the alerts will contain observation metadata (observation time, filter, etc.), astrometric and photometric measurements of the alerting source, measurements from earlier observations, and image cutouts of both the reference images and the science images. The alerts are not served directly to the scientific community, but are instead served to community transient brokers which are designed to filter, add value, and store the alerts \citep{graham_2024_11062350}. Several brokers exist, e.g., AleRCE \citep{2021AJ....161..141S}, Fink \citep{2021MNRAS.501.3272M}; with each focusing on a specific area of study, approach to processing alerts or adding specific information. Lasair \citep{2019RNAAS...3...26S,10.1093/rasti/rzae024} is the LSST:UK community broker, which is capable of processing incoming alerts rapidly in order to allow users to identify and determine the value of incoming alerts for follow-up, by allowing users to build bespoke cross matching filters. One of the key ways that Lasair does this is with Sherlock \citep{david_young_2023_8289325}, a contextual classifier which cross matches a transient against a library of catalogues in order to determine the most likely transient type, e.g., variable star, nuclear transient, supernova, etc.. Lasair is available to users through a website\footnote{\href{https://lasair-ztf.lsst.ac.uk}{https://lasair-ztf.lsst.ac.uk}}, an application programming interface (API), and pushed notifications of alerts. Currently, Lasair operates using Zwicky Transient Factory \citep[ZTF,][]{2019PASP..131a8002B} alerts, which has allowed Lasair to be robustly tested and developed before the start of Rubin operations, as the alerts are structured similarly.

\subsection{Combining LSST and \Euclid information for transients}
\citet{2022zndo...5836022G} proposed joint data products known as Rubin-Euclid Derived Data Products (DDPs), which will be created using proprietary Rubin and \Euclid science data processed in such a way as to protect the ``proprietary science'' of each survey. The DDP proposals contained several numbered products for transient science (DDP 25-35) with a specific urgency assigned to producing transient cutouts, light-curves, and a joint database and interface to access these. 

{There are several scientific benefits to combining \Euclid and Rubin observations of transients; the superior resolution provided by \Euclid allows for improved morphology measurements of a host galaxy, and localisation of the transient within the host. Similarly, \Euclid's resolution allows for the identification of compact hosts, whilst the NIR coverage allows for reddened hosts to be identified. \Euclid is also capable of identifying a huge number of galaxy lenses \citep[see e.g.][]{Q1-SP048}. Combining this ability with Rubin transient alerts will allow for a large number of galaxy lensed transients to be readily identified.}

\citet{Q1-SP002} carried out manual matching of archival known transients to \Euclid images from the first structured \Euclid data release \citep[Q1;][]{Q1-TP001}, i.e. the matching was not carried out in real time. However, in the Rubin era, with such an increase in the number of alerts per night, and with a desire to react quickly in case follow-up observations are required from other telescopes, we need an automated system.

There are two regimes in which transient DDPs can be created, one where the transient is first detected by \Euclid, and the other where it is first detected by Rubin. In this work we focus on the second regime, and present a prototype system for matching alerts received from Lasair to \Euclid observations. In Sect. \ref{sec:strategy} we describe the strategy used to match these alerts, and discuss how it would be impacted and enhanced by the establishment of a \Euclid-discovered transient stream. Section \ref{sec:implementation} outlines the technical implementation of this system, highlighting any specific choices that we have made. We describe the deployment of the prototype system, with particular focus on the user interface in Sect.\,\ref{sec:deployment}. Section \ref{sec:examples} discusses the performance of the system under prototyping operations and shows examples of matched alerts. Finally, in Sect. \ref{sec:nextSteps} we discuss the future direction of this work, including {predictions for rates of matched transients, and} how it will fold into an integrated transient DDP production pipeline.

\section{Transient Matching Strategy}\label{sec:strategy}
Matching transients from Rubin to \Euclid requires that alerts are matched for both spatial and temporal coincidence. Whilst spatial coincidence is a well-defined condition, the criterion for temporal coincidence is driven by the science goals. For example, transients such as superluminous SNe \citep{2018SSRv..214...59M} or pair instability SNe \citep{2024arXiv240716113R} have decay times of order 1 year, whereas type I/II SN are substantially shorter lived, but markedly more frequent \citep{2007ApJ...662..487W,2009ApJ...699.2026R,2021Natur.589...29B,1985AJ.....90.2303D}. As such, there is a challenge in selecting an appropriate temporal overlap that encompasses the diversity of transients without being na\"ive to the parameter space to which \Euclid and Rubin are sensitive.

\begin{figure}
\centering
\begin{tikzpicture}[node distance=1.05cm]
\node (88dd10f0-a3ee-4c37-8698-bb80a8d017dc) [7f27e395-8473-5b8a-b568-31425bce482a] {Rubin alert};
\node (e448dfac-a5b8-484f-a6a6-d4f2ec29d861) [69bbb168-da59-5865-902f-94e77902bf95, below of=88dd10f0-a3ee-4c37-8698-bb80a8d017dc, yshift=-0.1cm] {Crossmatch \Euclid Alerts};
\node (6aa237f9-fe3b-404c-b3b7-2a948fdc0972) [b0e005f8-e267-5a3a-a638-881fb9faed1d, below of=e448dfac-a5b8-484f-a6a6-d4f2ec29d861, yshift=-1.1cm] {Previous Alert?};
\node (Subsequent) [b0e005f8-e267-5a3a-a638-881fb9faed1d, right of=6aa237f9-fe3b-404c-b3b7-2a948fdc0972, xshift=2.1cm, yshift=-0.0cm] {Other Observations?};
\node (Observations) [b0e005f8-e267-5a3a-a638-881fb9faed1d, below of=6aa237f9-fe3b-404c-b3b7-2a948fdc0972, xshift=-0.0cm, yshift=-1.8cm] {\Euclid Observations?};
\node (NoAction) [512bdd77-c3aa-5669-a956-85f7a90c6fb4, left of=Observations, xshift=-1.55cm, yshift=-0.0cm] {No Action};
\node (f9624a31-43e7-4bb9-a699-54faa72c5a2d) [b4eeb85a-e852-5337-aaeb-b1b7ed281176, below of=Observations, xshift=-0.0cm, yshift=-1.2cm] {Forced Photometry};
\node (Combine) [b4eeb85a-e852-5337-aaeb-b1b7ed281176, below of=6aa237f9-fe3b-404c-b3b7-2a948fdc0972, xshift=-0.0cm, yshift=-5.5cm] {Combine};
\node (809d5e60-ab79-4849-bc8a-241fba129076) [574b93ce-5d31-5729-9ed0-9a79e73d4725, below of=Subsequent, xshift=0.0cm, yshift=-1.8cm] {Difference Image};
\node (Detection) [b0e005f8-e267-5a3a-a638-881fb9faed1d, below of=809d5e60-ab79-4849-bc8a-241fba129076, xshift=0.0cm, yshift=-1.2cm] {Detection?};
\node (5b0051da-629f-4ef8-a91c-3bb0340ff2b5) [512bdd77-c3aa-5669-a956-85f7a90c6fb4, below of=6aa237f9-fe3b-404c-b3b7-2a948fdc0972, yshift=-7.0cm, xshift=0.0cm] {Publish};

\draw [7be24b85-97d0-5b76-ba9e-d94005dca8f2] (88dd10f0-a3ee-4c37-8698-bb80a8d017dc) --  (e448dfac-a5b8-484f-a6a6-d4f2ec29d861);
\draw [7be24b85-97d0-5b76-ba9e-d94005dca8f2] (6aa237f9-fe3b-404c-b3b7-2a948fdc0972) --   node[anchor=south] {Yes} (Subsequent);
\draw [7be24b85-97d0-5b76-ba9e-d94005dca8f2] (e448dfac-a5b8-484f-a6a6-d4f2ec29d861) --  (6aa237f9-fe3b-404c-b3b7-2a948fdc0972);
\draw [7be24b85-97d0-5b76-ba9e-d94005dca8f2] (6aa237f9-fe3b-404c-b3b7-2a948fdc0972) --   node[anchor=east] {No} (Observations);
\draw [7be24b85-97d0-5b76-ba9e-d94005dca8f2] (Observations) --   node[anchor=south] {No} (NoAction);
\draw [7be24b85-97d0-5b76-ba9e-d94005dca8f2] (Observations) --   node[anchor=east] {Single} (f9624a31-43e7-4bb9-a699-54faa72c5a2d);
\draw [7be24b85-97d0-5b76-ba9e-d94005dca8f2] (Observations) --   node[anchor=south, xshift=-0.1cm] {Multiple} (809d5e60-ab79-4849-bc8a-241fba129076);
\draw [7be24b85-97d0-5b76-ba9e-d94005dca8f2] (Subsequent) -|   ([xshift=0.3cm, yshift=0cm]809d5e60-ab79-4849-bc8a-241fba129076.east) node[anchor=west] {No} |- (Combine) ;
\draw [7be24b85-97d0-5b76-ba9e-d94005dca8f2] (Subsequent) --   node[anchor=west, yshift=0.0cm,xshift=-0.0cm, align=center] {Yes} (809d5e60-ab79-4849-bc8a-241fba129076);
\draw [7be24b85-97d0-5b76-ba9e-d94005dca8f2] (809d5e60-ab79-4849-bc8a-241fba129076) --  (Detection);
\draw [7be24b85-97d0-5b76-ba9e-d94005dca8f2] (f9624a31-43e7-4bb9-a699-54faa72c5a2d) --  (Combine);
\draw [7be24b85-97d0-5b76-ba9e-d94005dca8f2] (Detection) --  node[anchor=south] {No} (f9624a31-43e7-4bb9-a699-54faa72c5a2d);
\draw [7be24b85-97d0-5b76-ba9e-d94005dca8f2] (Detection) |- node[anchor=south, yshift=0.2cm,xshift=-0.4cm] {Yes} (Combine);
\draw [7be24b85-97d0-5b76-ba9e-d94005dca8f2] (Combine) --  (5b0051da-629f-4ef8-a91c-3bb0340ff2b5);
\end{tikzpicture}
\caption{Schematic flow of operations in creating DDP following a Rubin alert. This shows the flow of operations in the situation where there exists \Euclid transient alerting infrastructure. Steps and decisions on the right hand side which require \Euclid transient alerts are excluded from this work. Nevertheless, they are shown here for completeness.}\label{fig:flowchart}
\end{figure}
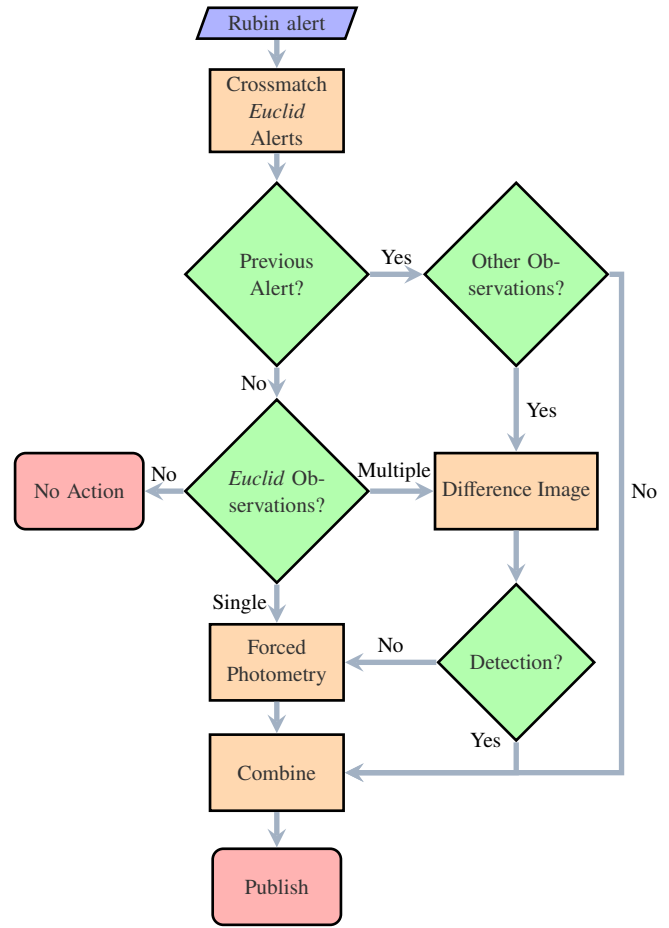

In developing a strategy for matching Rubin transients to \Euclid observations we remain mindful that a future \Euclid transient stream is envisioned, which should be integrated into any cross-matching strategy. Whilst this adds a degree of complexity and requires coordination, it brings the benefit of reducing duplicate analysis and can make the products produced here more data rich. \autoref{fig:flowchart} shows a schematic of the decisions and steps required going from a Rubin transient alert through to a final product, with steps dependent on the \Euclid transient detection pipeline shown on the right of the figure.

The simplest, and probably most common, case is where there has been no previous \Euclid transient alert coincident with the Rubin alert. Here, any \Euclid observation would be subject to single epoch forced photometry (i.e. a photometric measurement at the position of the transient provided by Rubin) and the data combined with the Rubin data before being published. Where more than one epoch of \Euclid observations exist, then these transient coordinates should be passed to the \Euclid transient detection pipeline for source detection and processing; where no source is detected each epoch should then be subject to forced photometry. In the event that there has been a \Euclid alert and there have been \Euclid observations either subsequent or prior to a \Euclid alert, those observations should also be processed by the \Euclid transient detection pipeline, where again should no source be identified in each of those observations, then those observations should be subject to single epoch forced photometry. Finally, where there has been a previous \Euclid transient alert and no further \Euclid observations, both sets of transient alerts should be combined and published.

Until such time as a full \Euclid transient stream is operational it is of course not possible to include information from it in the strategy implemented here. This is of limited impact for the work which we present here as matching does not rely on the preexistence of a \Euclid transient (the vast majority of Rubin transients will not have been previously detected by the \Euclid transient stream), and at most this means that information which \textit{could} be provided by a \Euclid transient stream is not. However, the observations of a \Euclid transient which would have formed part of an alert stream will continue to be considered regardless. The steps in Fig.~\ref{fig:flowchart} that are dependent on the existence of a preexisting \Euclid alert are excluded from the matching strategy as implemented in this work. Currently in the absence of a \Euclid transient alert stream, we have a simple flow to follow; we need only confirm if there have been \Euclid observations with a suitable temporal overlap with the Rubin alert.

\section{Implementation}\label{sec:implementation}

Implementing this strategy in an accurate, scalable, and resource efficient way is a multi-step process which leverages pre-existing technologies and uses algorithms developed as part of this work.

In the following we describe a prototype transient matching system. We are using real \Euclid data (see Sect.~\ref{sec:Euclid}), and to mimic Rubin alerts, we use the Lasair alert stream, which is currently processing alerts from the Zwicky Transient Facility (ZTF). Briefly, ZTF is a survey running on the 48 inch Palomar telescope with a $\sim47\,{\rm deg}^{2}$ field of view that surveys the northern sky at a cadence of approximately two days. ZTF publicly releases photometry in the Sloan Digital Sky Survey (SDSS) \textit{g}- and \textit{r}-bands to a limiting magnitude of $\sim$20.5--21. As ZTF does not have the same photometric depth or range of filters as Rubin it should be noted that the types and rates of alerts will differ.

\subsection{Lasair alert ingestion}
Using Lasair, the ingestion of relevant alerts is a relatively simple process, as Lasair allows for the generation of user-defined alert filters which can be used to select alerts to be streamed directly to the user using a \texttt{Kafka}\footnote{\texttt{Apache Kafka} is a platform for processing and handling large data streams: \href{https://kafka.apache.org}{kafka.apache.org}} stream. For the purposes of this work we have streamed alerts that have their date of maximum brightness within 100 days of the current date and those which have not been identified by \texttt{sherlock} as either a variable star, bright star, cataclysmic variable, or active galactic nuclei (AGN). This helps to limit the number of alerts which we have to consider to those that are most likely to have a useful \Euclid observation from the standpoint of transient science, {and reduce the number of bogus alerts ingested}. It is further possible to limit the alert stream to areas of the sky that have a desired spatial and temporal overlap with \Euclid, e.g., by limiting the considered sky area to the current year of planned \Euclid observations. However, at this time we only limit the stream to the area of the EWS footprint.

Information considered relevant from the alerts that are streamed from Lasair via a \texttt{Kafka} stream are saved in a mySQL database table, \texttt{incomingAlerts}. This table's schema is shown in \autoref{tab:incomingAlerts} and contains the key information about the streamed alert. \texttt{gmag} and \texttt{rmag} are allowed to take null values because many alerts only trigger in one band, \texttt{lastConsidered} is also allowed to take a null value because it is not part of the streamed alert and is a table column which is used to track when a given alert was most recently considered for matching \Euclid observations (see Sect. \ref{sec:crossmatching}). Each entry is uniquely tracked with the \texttt{objectId} which takes a unique value identical to ZTF object IDs, thus in cases where a second Lasair alert is issued for the same object, the entry is updated with new details. {When operations include Rubin alerts the table schema will be expanded to include table columns for the filters which are present in Rubin but not ZTF.}

\begin{table}
    \centering
    \caption{\texttt{incomingAlerts} table schema. Field names, with the exception of \texttt{lastConsidered}, are taken from the Lasair alert stream, or a contraction thereof. For the sake clarity \texttt{jdmax} is the jullian day of the brightest point in the Lasair lightcurve and \texttt{utc} is the time that the ingested alert was triggered. These dates are not necessarily the same.}
    \label{tab:incomingAlerts}
    \begin{tabular}{ccc}
        \hline
        \hline
        Field & Type & Null  \\
        \hline
        \texttt{objectId} & varchar(20) & NO   \\
        \texttt{gmag} & float & YES     \\
        \texttt{rmag} & float & YES    \\
        \texttt{jdmax} & float & NO    \\
        \texttt{ramean} & float & NO    \\ 
        \texttt{decmean} & float & NO    \\ 
        \texttt{utc} & datetime & YES    \\
        \texttt{sherlockClass} & varchar(100) & YES    \\ 
        \texttt{lastConsidered} & datetime & YES    \\
        \hline
        \hline
    \end{tabular}

\end{table}

\subsection{Choosing alerts to cross-match with \Euclid}
Whilst it might be tempting to check every alert all of the time, this is neither feasible nor efficient and thus it is necessary to use some selection criteria. In order to do this we use the \texttt{lastConsidered} column to join \texttt{incomingAlerts} to the table \texttt{checkList} (see \autoref{tab:checkList}), which contains all of the previously matched alerts. Joining these two tables allows us to select alerts which meet any of the following criteria\footnote{Whilst this replicates the logic of the \texttt{sql} this is not the exact form of the \texttt{WHERE} clause used.}:
\begin{itemize}
    \item the alert has never been considered (e.g. $\rm lastConsidered == \rm Null$);
    \item the alert has previously been matched and it has not been checked for more than 28\,d (e.g. $\rm EuclidObservationDate != \rm Null \wedge \rm lastConsidered < \rm TODAY - 28\,d$);
    \item the alert has previously been considered but not matched and it has not been checked for more than 7\,d and the alert was triggered less than 120\,d ago (e.g. $ \rm EuclidObservationDate == \rm Null \wedge \rm lastConsidered < \rm TODAY - 7\,d \wedge \rm utc > \rm TODAY - 120\,d$\footnote{120\,d was chosen as it is twice the temporal overlap we use in this work, see Sect. \ref{sec:temporal}}).
\end{itemize}

By using these criteria, we can ensure that all alerts are considered and that alerts are considered sufficiently often to ensure that they are matched timeously without being an undue burden on resources. Setting an upper limit on the age of an alert further aids in resource management by eliminating events that will never have \Euclid observations within the desired overlap for the automatic matching program. 

\begin{table}
    \centering
    \caption{\texttt{checkList} table schema. Field names match those given in \autoref{tab:incomingAlerts} which the exception of \texttt{ZTFDate} which is \texttt{utc} renamed in order to be more descriptive.}
    \label{tab:checkList}
    \begin{tabular}{ccc}
        \hline
        \hline
        Field & Type & Null \\
        \hline
        \texttt{objectId} & varchar(20) & NO   \\
        \texttt{gmag} & float & YES     \\
        \texttt{rmag} & float & YES     \\
        \texttt{jdmax} & float & NO     \\
        \texttt{ramean} & float & NO    \\
        \texttt{decmean} & float & NO     \\
        \texttt{alertDate} & datetime & YES     \\
        \texttt{cutoutPath} & varchar(600) & YES     \\
        \texttt{ZTFDate} & datetime & YES     \\
        \texttt{sherlockClass} & varchar(100) & YES     \\
        \texttt{EuclidObservationDate} & datetime & YES  \\
        \texttt{lastConsidered} & datetime & YES   \\
        \texttt{lowResPath} & varchar(250) & YES   \\
        \texttt{tolerance} & int & YES  \\
        \texttt{interestFlag} & varchar(8) & YES  \\
        \hline
        \hline
    \end{tabular}
\end{table}

\subsection{\Euclid lookup and cross-matching}\label{sec:crossmatching}
All \Euclid image products, catalogues, and ancillary tables are stored and accessed through the Euclid Science Archive\footnote{\href{https://eas.esac.esa.int/sas/}{eas.esac.esa.int/sas/}, note that this is the publicly accessible version of the Euclid Science Archive, though functionally the same this work makes use of an internal version of the archive.}. The primary \Euclid imaging data products in all 4 bands are known as Level 2 Mosaics. These are image products that are arranged as predefined, regularly-shaped tiles of the sky, {with pixel scales of of 0\arcsecf1 for $I_{\scriptscriptstyle\rm{E}}$ and 0\arcsecf3 NIR observations,} and are composed of different observations \citep[see Sect. 7][for a full description]{2022A&A...662A.112E}. As a direct consequence of this construction, these products are lacking precise time information, and provide only a range from the first to last constituent observation, which can span days; as such we make use of an intermediate data product known as Level 2 Stacked Image (hereafter LE2 stacks). These are processed (flat fielded, cosmic ray removal, world coordinate system (WCS) computed) image products composed of the combination of 4 dithered consecutive exposures which constitute a single epoch with a field of view of $0.53\,\rm deg^{2}$. The stacking of four dithers serves to increase depth, eliminate chip gaps and reduce the effect of cosmic ray strikes \citep[a complete description can be found in Sect. 3 \& 4 of][]{2022A&A...662A.112E}. 

We search for LE2 stacks that spatially overlap the Lasair alert within 30\arcsec\footnote{This large radius is chosen because the image products we are creating are of this size and some science cases such as host identification benefit from partial overlap.} by querying the \texttt{sedm.observation\_stack} table in the \Euclid ``On The Fly'' database (OTF) via TAP\footnote{Table Access Protocol is an IVOA (International Virtual Observatory Alliance) standard for accessing an ADQL database remotely.} upload and ADQL query. In order to manage resources and minimise the effect of a dropped connection to the OTF database, a maximum of 100\,000 alerts are checked at any one time and OTF queries are limited to 10\,000 alerts per job. Following the OTF database query, if an alert has no spatial overlap then the \texttt{lastConsidered} column corresponding to its entry in \texttt{incomingAlerts} is set to the current date and time so that the alert can continue to be regularly checked for spatial overlap in observations and ultimately disregarded if none exist within the 120 days specified.

\subsubsection{Temporal matching}\label{sec:temporal}
Having identified those alerts that have spatial coincidence, we must identify those with an appropriate temporal coincidence. For the purposes of this pilot work we have defined that temporal coincidence criterion as $\pm60\,\rm d$, i.e. \Euclid observed at the location of the alert within $60\,\rm d$ before or after the alert date. This temporal criterion was chosen mindful of the classes of transient generally observed by ZTF; in operations with Rubin we will consider using a less restrictive criterion. 

In order to identify those alerts that meet the temporal criterion with respect to \Euclid observations, we check each \Euclid observation which is spatially coincident with an alert and identify for each \Euclid band which, if any, observations meet the defined temporal criterion.

Confirming that a \Euclid observation meets the temporal coincidence criterion is straightforward where there is only one observation, as will be the case in the EWS, as we need only determine if the observation meets the defined temporal criterion. Where there are more than one \Euclid observation, e.g., in the EDFs, we check each observation to confirm which, if any, meet the temporal criterion. In the rare case where more than one observation meets the temporal criterion we select the observation that took place closest in time to the alert. This choice is made as the closest observations in time are those most likely to contain the transient and thus allow for easy visual confirmation by a user.

The now matched alert is then added to the \texttt{checkList} table with the date and time of the selected observations added to the \texttt{EuclidObservationDate} column and \texttt{alertDate} and \texttt{lastConsidered} set to the current date if this is the first time this alert has been considered. If it is not the first time it has been considered, then only \texttt{lastConsidered} is updated.

\subsection{Gathering observations}
Until this point in the implementation we have only considered observations as they are recorded in databases or alert streams. In order to perform science, we must gather the selected observations. Gathering data from Lasair is a straightforward process due to the functionalities provided by the \texttt{lasair} python module which allows us to directly query Lasair for all the information held on a given object. This means that the provided image postage stamps (reference, science, and difference) for each epoch can be readily accessed to serve to a user, alongside the photometry.

\Euclid observations are gathered using the \texttt{requests} python module which allows us to download directly from the \Euclid cutout server via HTTP\footnote{The Euclid Science Archive is not currently fully compliant with IVOA standards; this is a planned feature (Altieri, private communication). When this happens we would transition this to a VO implementation.}, using file paths returned by the ADQL query to the OTF outlined in Sect. \ref{sec:crossmatching}. These allow us to collect $30\arcsec\times30\arcsec$ cutouts in each of the bands, centred on the target location from the previously identified observations. Unlike Lasair, there is no \Euclid product which provides photometry based on the LE2 stacks\footnote{The \Euclid mosaic catalogue products do provide Kron and image segmentation photometry of detected objects. Although these provide some temporal information, as previously discussed are not suitable for our purpose.}. As such it is necessary for us to compute this ourselves.

\subsubsection{Photometry}\label{sec:photometry}

In this prototype we have assumed that all \Euclid observations are single epoch and we have implemented, for development purposes, a na\"ive approach to photometry; doing forced aperture photometry at the location of the alert. This has the advantage that it is computationally cheap and reliable for isolated sources making it suitable for development work to demonstrate functionality. However, it is not well suited to returning scientifically robust results, particularly for extra-galactic transients where their host galaxies' ``local background'' can contaminate measurements. Transient surveys usually overcome this issue by relying on difference images, but with \Euclid that is generally not an option outside of the deep fields, which provide multi-epoch observations. 

As development has matured we have worked to improve the robustness of the photometry by implementing forced PSF (point spread function) photometry using \texttt{photutils} \citep{larry_bradley_2025_14606896}, using PSF models provided by \Euclid. This requires that we access a PSF model from the Euclid Science Archive for each filter and each pointing. These are small files that we currently access on an as needed basis, though we are considering the benefits and viability of pre-emptively storing these and accessing them via a lookup table when needed. Although PSF photometry can still be impacted by contaminating light it can provide far more robust results. It also allows us to estimate if there has been source detection or not by using the in-built quality-of-fit metric\footnote{Defined as the absolute value of the sum of the fit residuals divided by the fit flux.} and allow us to report upper limits as needed. Not all elements of this have been fully implemented to the deployment (Sect. \ref{sec:deployment}), such as robust upper limits for especially faint sources, but nevertheless we demonstrate the photometry and the improvement offered by the PSF method in Sect. \ref{sec:examples}.

In the regime of multi-epoch observations, which is outside the scope of this work (see however Sect.~\ref{sec:matcheddetails}), photometry will be undertaken in the first instance via the transient detection pipeline, which will utilise difference imaging for source detection. This will provide, where a source is detected, photometry which is not impacted by the effects of ``local background''. Where a source is not detected it can be subjected to the forced photometry procedure which can supply upper limits. {This will also allows for more robust error estimation to be undertaken by comparing photometry from image subtraction with PSF and aperture photometry. Future implementation of photometry may also include deblending procedures for particularly crowded fields; tools such as \texttt{SCARLET} and \texttt{SCARLET 2} \citep{2018A&C....24..129M,2024A&C....4900875S} already exist to do this and could be included.}

\subsection{Catch up steps}
Due to the data processing implementation of the \Euclid mission, where images taken with two different instruments are reduced using separate pipelines, it can be the case that data from one instrument is available to match with an alert but not the other. When this happens, knowing that data from the other instrument will become available, we run a frequent, targetted search for this ``missing'' data. This search is not limited by when it was last executed and is run on alerts where one or more \Euclid bands are missing from the gathered observations. In this respect the search is not matching data to alerts, it is only attempting to locate data that is expected to be in the OTF. When ``missing'' data is found, it is downloaded and the database is updated accordingly. This is a resource-efficient way to ensure the completeness of the matched records as the number of calls is relatively small, thus we can run the method frequently on a per alert basis which would not be feasible for the complete alert stream. We expect that in steady state operations the time gap between data from instruments becoming available will be small and the number of executions of this search will be small.

\section{Deployment}\label{sec:deployment}
In the preceding sections we described the steps we undertook to build a joint product from a Rubin alert with \Euclid observations. Here we describe the practical steps we take to deploy this body of work to serve usable products to users. This can be split up into two aspects; a back end that monitors the alert stream, populates the database, and gathers observations; and a front end which serves the product to the user. We manage both aspects by using the Spin system at National Energy Research Scientific Computing Center (NERSC)\footnote{\href{https://www.nersc.gov/what-we-do/computing-for-science/data-resources/spin}{https://www.nersc.gov/what-we-do/computing-for-science/data-resources/spin}} which allows us to deploy docker containers to run both the front and back end. 

\subsection{Back end}
The back end elements of the system are executed every 6 hours controlled by a cron job. This runs the entirety of the cross matching processes described above in addition to other steps which support the front end; for example generating image preview cutouts to be shown as part of an alert preview. We selected 6 hours as a repeat schedule in order to balance the needs of frequent update with the time that it can take a large batch of matching to complete. We find that most matching runs complete on the order of minutes but when there are of order 100\,000 alerts to match (which is plausible under LSST operations) this can reach several hours. Choosing a schedule that ensures that only one run is occurring at a time prevents a duplication of efforts or unexpected database entries. {In anticipation of the possibility that the number of Rubin alerts exceeds this schedule and is prone to inducing delay, duplication, and database errors the batching of alerts and queries discussed in Sect.~\ref{sec:crossmatching} can be parallelised if needed, handing off different batches to separate workers, with minimal update to the codebase.}

\subsection{Front end}
The front end is managed using the \texttt{Django} web framework\footnote{\href{https://www.djangoproject.com}{https://www.djangoproject.com}}, as this allows us to build a web front end that can easily interact with the database and scale with the number of matched alerts. Within the front end there are three main aspects which we have built. Although it is not necessary to discuss here, it should be noted that due to the proprietary restrictions placed on firstly \Euclid data and secondly Rubin-\Euclid DDPs, the front end is limited to authenticated users. 

\subsubsection{Home Page/Summary Page}

The primary (and first page that an authenticated user sees) is the home page, which contains a summary table of the most recently matched alerts. For each matched object, this shows the alert name, coordinates, the dates of both the Lasair alert and the \Euclid observation that most closely matches in time, the Sherlock classification and a $5\arcsec\times5\arcsec$, \Euclid $I_{\scriptscriptstyle\rm{E}}$ image cutout. Additionally the table entries highlighted in {green} indicate sources where at least one of the \Euclid detections of the source is brighter than $24$ mag, indicating that it may be especially worthy of consideration as it is within the brightness range of many follow-up instruments. The \texttt{interestFlag} column is used to store a flag to denote such sources, and is added by running the photometric procedure detailed in Sect. \ref{sec:photometry}.

Additionally, the table has a number of interactive elements which makes it fully searchable. This is particularly useful for locating a specific source or limiting to a given Sherlock classification. Furthermore, it is possible to use the slider at the top of the page to restrict the temporal separation between \Euclid and Rubin observations, with the default to show everything that has been matched.

\subsubsection{Matched details}\label{sec:matcheddetails}
\begin{figure}
    \centering
    \fbox{\includegraphics[width=\linewidth]{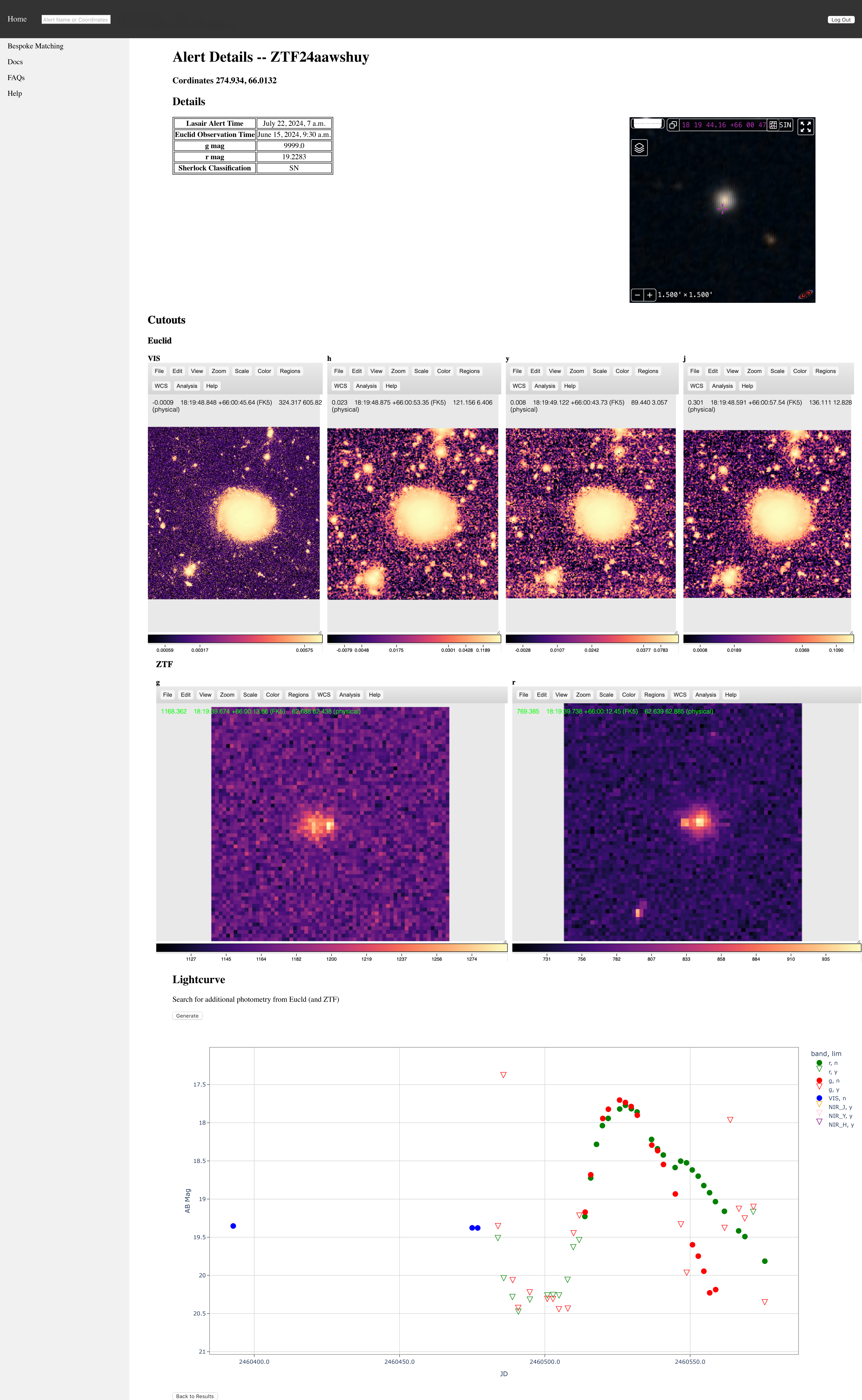}}
    \caption{Example of a joint alert page showing ZTF24aawshuy.}
    \label{fig:alert-details}
\end{figure}

More details on a source can be seen in the detailed view for any of the matched sources, e.g., Fig.~\ref{fig:alert-details}. This page contains several useful elements to understand the transient and its context. The top of the page contains information from the Lasair alert including the date of alert and the brightness in the different observation bands used in the alert. Beside this is an embedded interactive Aladin Lite\footnote{\href{http://aladin.cds.unistra.fr/AladinLite/}{http://aladin.cds.unistra.fr/AladinLite/}} frame centred on the object. By default this shows the object in a $1\arcminf5\times1\arcminf5$ cutout from SDSS, but this can be easily changed by the user to a scale and survey that they find useful for the object they are considering.

In the middle of the page there is an array of JS9\footnote{\href{https://js9.si.edu}{https://js9.si.edu}} windows which display the science images from each of the \Euclid and ZTF bands. Where an image isn't available, a placeholder informs the user of this. JS9 is a JavaScript based successor to DS9 used to explore and inspect FITS images. Using JS9 allows us to align the WCS of the images and synchronise zooming and panning operations as well and allows users to alter image scaling, brightness cuts and to inspect specific elements of the images; e.g., the 2-D profiles.

At the bottom of the page is an interactive plot which shows the joint lightcurve combining the photometry provided by Lasair with \Euclid photometry which we calculate as described in Sect. \ref{sec:photometry}. This plot shows both the brightness of object detections and in their absence the limiting magnitude of the observations; in ZTF this is a variable value whereas in \Euclid, when using aperture photometry, this is an assumed fixed value for the VIS and NISP instruments as given in \citet{EuclidSkyVIS} and \citet{EuclidSkyNISP} respectively (although in reality, these  limits will vary slightly by field). Going forward, as we implement PSF photometry, we will use the resultant statistics, as previously discussed, to identify upper limits in \Euclid data. Currently, we do not store the photometry and it is calculated as needed when matched objects are inspected; this is because the photometry procedure is computationally lightweight enough to be done during other necessary loading steps. As such we can save on the amount of storage required by not storing photometry for alerts that may be discounted by users directly from the summary page. Implementation of PSF photometry as detailed in Sect. \ref{sec:photometry} is unlikely to change this as the process is not perceptively slower than the aperture photometry that was in use when this aspect of the system was designed.

Since the \Euclid transient pipeline is yet to be integrated with our system, we act as if every \Euclid observation is single epoch, which will be the vast majority of cases. As such we don't currently store any \Euclid images from other epochs that may exist. In order to get complete \Euclid photometry for those objects that do have multiple epochs, we include an option to generate the complete \Euclid light-curve. This accesses all \Euclid observations for the object, performs photometry, and adds these data points to the light-curve plot. It has not yet been decided how the full treatment of multi-epoch photometry is to be implemented; in part this decision will be guided by the implementation of the \Euclid transient detection pipeline. It may be the case that all photometry for these objects specifically is stored as it may not be viable to calculate difference imaging on an as needed basis.

\subsubsection{User defined matching}\label{sec:userDefined}
As we discussed in Sect. \ref{sec:strategy} \& \ref{sec:temporal}, alerts are matched against a fixed temporal overlap, but this is not necessarily ideal for all use cases, e.g., for pair instability SN where far greater separation in time may still result in a useful match. In order to allow for this we also provide the ability for users to perform a query of their own defined temporal overlap on either a known source or a cone search. This search function accesses the functions used by the back end to search Lasair for transient alerts before querying \Euclid using the overlap parameters provided by the user.

A summary page similar to that of the home page is returned by the search which contains all of the matched alerts. This allows the user to select those which they are interested in inspecting further. Doing so takes them into a detail page identical in layout to that of Fig.~\ref{fig:alert-details}.

\section{The system in use}\label{sec:examples}

While the system was operating over the period of July--December 2024, approximately 13\,000 Lasair alerts were matched to \Euclid observations. During much of this time, until mid-November 2024, \Euclid observations were focused on the southern sky which has limited spatial overlap with ZTF. Thus we expect, and indeed observed, a far higher rate of matched alerts at times when \Euclid is observing the sky area from which the alerts are coming. It should however be noted that for a substantial number of these ``matched'' alerts, \Euclid will not have detected the transient. This is simply because very early time observations by \Euclid are unlikely to detect the transient as the transient detonation will not yet have occurred. In this case it is possible however, that for some transients such as type II SNe, that we will be able to identify some precursor activity \citep[e.g.][]{Pastorello07,Ofek13,Jacobson-Galan22}. Additionally, approximately 75\% of the matches in the system are classified as orphans by Sherlock (transients without a known host). Despite this, \Euclid will be capable of adding scientific value for many of these, even if the transient is not observed, by e.g., detecting a possible host. There will however be cases that neither the host nor the transient are detected by \Euclid as the host will be too faint and the transient will either also be too faint or have detonated after \Euclid observations.

Matched alerts can fall into three categories: matches where \Euclid observed the transient but it was not possible to confirm a definitive detection but only upper limits, matches where \Euclid observed the transient and it was possible to confirm a definitive detection, and matches where the transient is absent entirely from \Euclid observations at its reported location. This final case is of course entirely agnostic of any temporal constraint and could in principle be applied to any Rubin transient. In the following we present some examples of these cases, with data extracted and redisplayed for this publication.

\subsection{\Euclid observed transients}
There are two cases where \Euclid observes the transient matched to a Lasair alert: where \Euclid observed before the alert and where \Euclid observed after the alert. This first case is interesting because it allows us to better constrain the detonation time of the transient and additionally adds further early time multicolour information; both of which can be useful in understanding the progenitor of the system. In the latter case this can be useful to understand the evolution of the lightcurve and help improve classification. Here we present one example of each that also appears in \citet{Q1-SP002}, which allows us to compare the quality of the product produced here through automatic matching with that of the manual approach undertaken in that work. The examples which we present here do have multiple epochs of observations; ordinarily as discussed in Sect. \ref{sec:strategy} these observations would have at least in part been processed through the \Euclid transient discovery pipeline, in its absence we treat each observation as if it were a single epoch observation. {Throughout this section we show both aperture and PSF photometry. We do this to highlight both success of the PSF photometry with the \Euclid provided PSF models and the shortcomings of using forced aperture photometry}

\subsubsection{ZTF24aawshuy}\label{sec:ZTF24aawshuy}
\begin{figure}
    \centering
    \includegraphics[width=\linewidth]{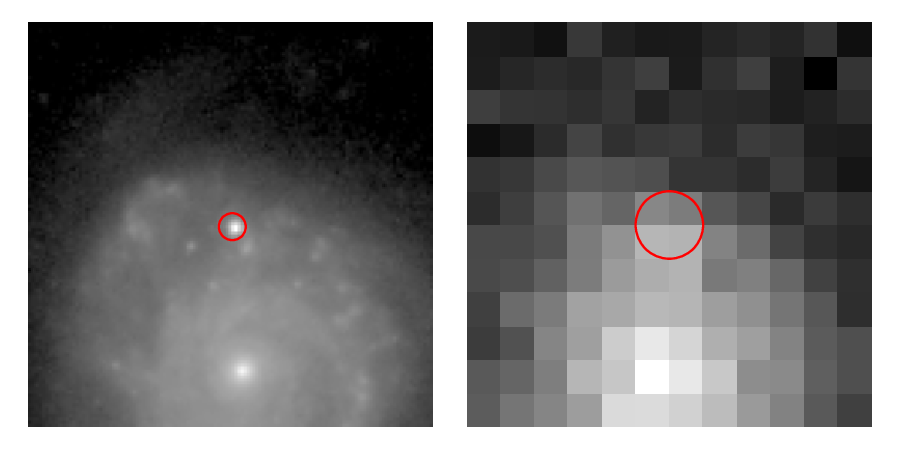}
    \includegraphics[width=\linewidth]{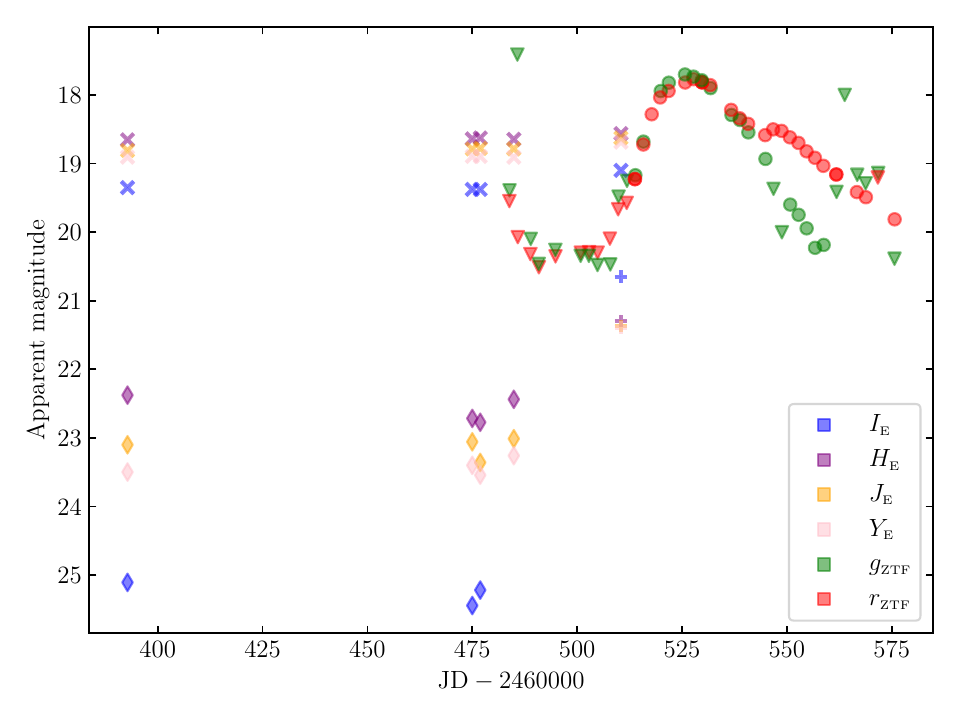}
    \caption{ZTF24aawshuy: \textit{Upper left:} \Euclid $I_{\scriptscriptstyle\rm{E}}$ band $12\arcsec\times12\arcsec$ cutout at $\rm JD=2460510$. \textit{Upper right:} ZTF \textit{r}-band $12\arcsec\times12\arcsec$ cutout of the detection image at $\rm JD=2460513$; in both North on top and East to the left. \textit{Lower:} Joint light curve, where detections are shown with filed circles and upper limits with triangles. Circles show and triangles show ZTF detections and upper limits respectively. Plus and diamonds show \Euclid PSF photometry detections and upper limits respectively. Crosses show \Euclid forced aperture photometry; these appear consistently brighter due to the local background contribution from the host galaxy.} 
    \label{fig:SN2024pvw}
\end{figure}

ZTF24aawshuy, also known as SN 2024pvw, is a known Type Ia SN \citep{2024TNSCR2607....1S} first discovered by ZTF on 2024-07-22. Preceding this there were full (VIS+NISP) sets of \Euclid observations on 2024-03-21, 2024-06-13, and 2024-06-15 with the final of these being the observations which were formally matched. In addition to this there were NISP-only observations on 2024-06-23. There was a further observation on 2024-07-18 which was part of the \Euclid Q1 data release \citep{Q1cite}, which was not made available through the OTF (though in steady state operations would have been). For completeness, we have manually ingested and processed these data through the same steps as if they had come from the OTF, though it should be noted that these data were provided in a different product, with different background subtractions used. In Fig.~\ref{fig:SN2024pvw} we show the $I_{\scriptscriptstyle\rm{E}}$ cutout from 2024-07-18 and ZTF \textit{r} cutout zoomed on the transient location and the lightcurve built by our system showing both approaches taken to photometry as detailed in Sect. \ref{sec:photometry}.

The cutouts in Fig.~\ref{fig:SN2024pvw} highlight the power of \Euclid; not only can the structure of the host galaxy be seen far more clearly than in ZTF due to the far greater resolution and depth, but the SN can be seen immediately by eye unlike in the ZTF image which relied upon differencing techniques. Although this difference will be substantially reduced when working with LSST data, it demonstrates that even with a single \Euclid exposure, \Euclid is capable of contributing meaningfully to transient science.

In the lower panel of Fig.~\ref{fig:SN2024pvw} we show both \Euclid PSF photometry (detections, $+$; upper limits $\smallblackdiamond$) and the na\"ive \Euclid forced aperture photometry ($\times$) addition to the ZTF photometry provided by Lasair. The shortcomings in the na\"ive approach of forced aperture photometry are readily apparent, with all of the \Euclid data exceeding the ZTF upper limits and host light contamination significantly degrading the measurements. This contamination is more pronounced in the NISP data due to the larger pixel scale compared to VIS. Although these aperture photometry measurements are well above the limiting magnitude for both instruments and despite the SN being readily identifiable by eye, it is impossible, due to the level of contamination, to call any of these detections. Using PSF photometry we recover far more plausible values. Similar to the findings of \citet{Q1-SP002}, we recover for each of the \Euclid bands $I_{\scriptscriptstyle\rm{E}}=20.69\pm0.041$, $Y_{\scriptscriptstyle\rm{E}}=20.73\pm0.034$, $J_{\scriptscriptstyle\rm{E}}=21.02\pm0.031$, $H_{\scriptscriptstyle\rm{E}}=21.19\pm0.028$ for the most recent \Euclid observation which corresponds to the first epoch after detonation. At all remaining epochs we recover only upper limits, though this should be expected as these observations were made substantially before detonation.

This lightcurve allows us to constrain the detonation time of the SN by providing reliable photometry from earlier in the rise phase post detonation. Additionally, the deeper upper limits that can be obtained from \Euclid prior to detonation can also aid in constraining the detonation date. This can be helpful in informing issues such as those pertaining to the nature of the secondary star in the white dwarf binary progenitor system.

PSF photometry, in the absence of difference photometry, is demonstrably the superior technique as it can mitigate for the effects of host contamination. Furthermore, as it gives quality of fit metrics it is possible to determine in a more nuanced way than just using limiting magnitudes if a source has been ``detected''. This comes with the caveat that it is impossible, without difference imaging, to definitively claim a transient detection as it could be, for example, a contaminating stellar source. In this specific case it would be possible to make this determination because multiple epochs of \Euclid observations exist. However, for the vast majority of targets this will not be the case.

\subsubsection{ZTF24aauueye}
\begin{figure}
    \centering
    \includegraphics[width=\linewidth]{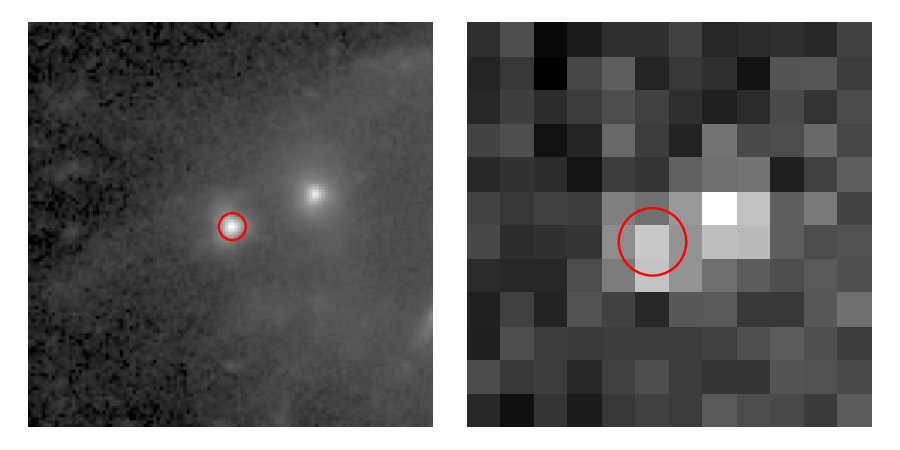}
    \includegraphics[width=\linewidth]{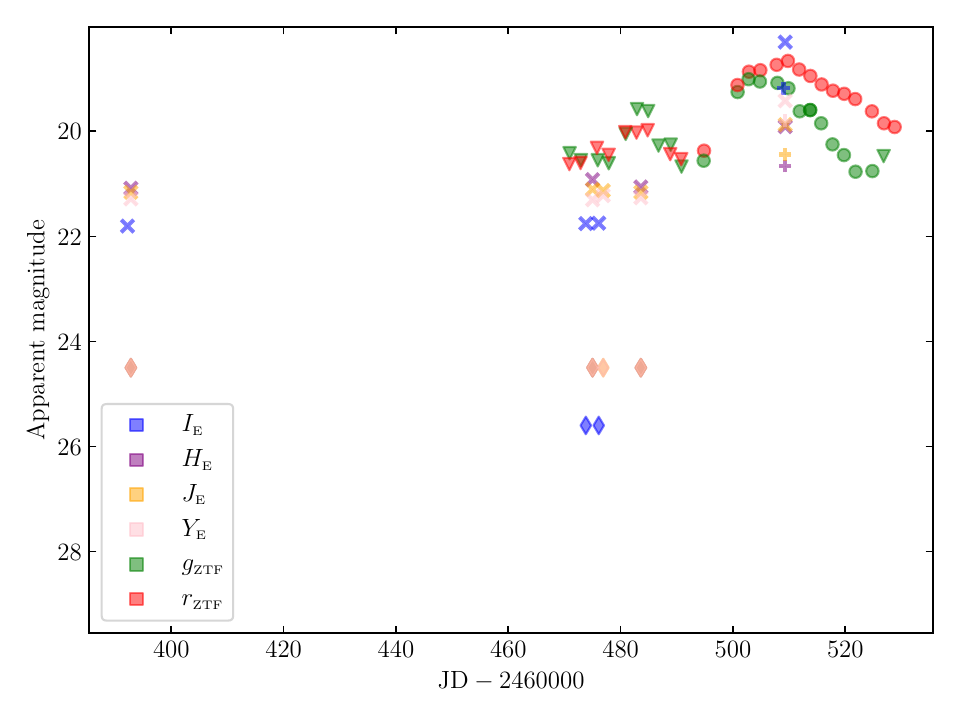}
    \caption{ZTF24aauueye. \textit{Top:} $30\arcsec\times30\arcsec$ \Euclid $I_{\scriptscriptstyle\rm{E}}$ band image at $\rm JD=2460509$ and ZTF \textit{g}-band discovery image at $\rm JD=2460494$; in both North on top and East to the left. \textit{Bottom:} Light curve as in Fig.~\ref{fig:SN2024pvw}.}%
    \label{fig:ZTF24aauueye}
\end{figure}

ZTF24aauueye (also known as AT 2024pcm) is a probable SN identified by ZTF \citep{Sollerman24}, though due to its relatively faint discovery brightness no spectroscopic classification is available. It was discovered on 2024-07-03 and \Euclid observed its location on 2024-03-24, 2024-06-12 and 2024-06-14. Similar to ZTF24aawshuy, above, there was an additional observation in the Q1 data release on 2024-07-18. In Fig.~\ref{fig:ZTF24aauueye} the \Euclid $I_{\scriptscriptstyle\rm{E}}$ cutout, ZTF \textit{g} discovery cutout, and the combined lightcurve are shown.

The quality of photometry, and the differences between the two methods are very similar to that presented in Sect. \ref{sec:ZTF24aawshuy} with the notable exception that there are no abnormally bright aperture photometry points exceeding the ground based measured limits. This is because ZTF24aauueye is subject to less host contamination than ZTF24aawshuy. The pre-detonation epochs for ZTF24aauueye are notably faint, and we do not recover magnitudes in any band above the quoted limiting magnitudes using PSF photometry in any band, as such for each of we plot the instrumental limiting magnitude at the appropriate epoch. In full deployment we would consider a more robust method of measuring limiting magnitude on a per epoch basis in under to provide more reliable measurements.

Comparing with \citet{Q1-SP002}, who report $J_{\scriptscriptstyle\rm{E}} = 20.284$ as their only measurement for this source, we demonstrate the ability to return measurements in all 4 \Euclid bands, as we are not reliant on constructing our PSF model (which was not always possible in that work). For the same epoch we measure $I_{\scriptscriptstyle\rm{E}} = 19.17\pm0.048$, $Y_{\scriptscriptstyle\rm{E}} = 19.797\pm0.023$ , $J_{\scriptscriptstyle\rm{E}} = 20.31\pm0.020$, and $H_{\scriptscriptstyle\rm{E}} = 20.47\pm0.019$ using the PSF photometry method. The comparable $J_{\scriptscriptstyle\rm{E}}$ measurement in addition to successfully measuring the other bands demonstrates the advantages of using pre-calculated PSF models for automatic photometry of matched alerts.

\subsection{No \Euclid detected transient}
As discussed previously, for many of the matched alerts we will not see the transient in \Euclid observations. These can however still be useful, for example by constraining the explosion epoch, by adding host information to sources, or to identify hosts for transients otherwise identified as orphans. Below we present examples of the latter two cases.

\subsubsection{ZTF24abxylal}
\begin{figure}
    \centering
    \includegraphics[width=\linewidth]{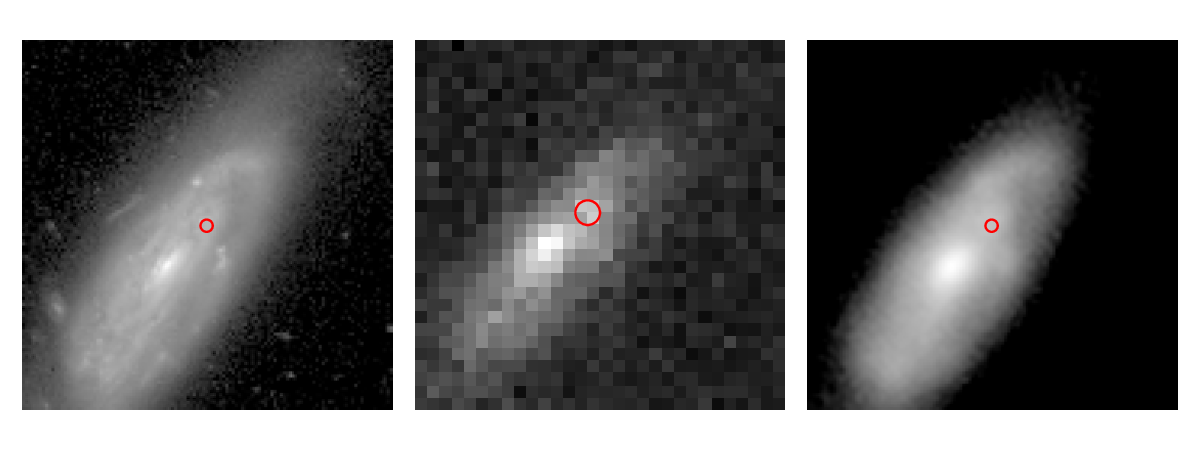}
        \caption{ZTF24abxylalc: $30\arcsec\times30\arcsec$ of \textit{left}: \Euclid $I_{\scriptscriptstyle\rm{E}}$ band, \textit{centre:} ZTF \textit{r}-band discovery image, \textit{right:} Legacy Survey \textit{g}-band; in all North on top and East to the left.}
    \label{fig:host}
\end{figure}

The location of ZTF24abxylal (SN 2024aebt) was observed by \Euclid 60 days prior to discovery by ZTF. The transient was subsequently classified as a Type II SN by \citet{2024TNSCR4992....1T}. The observation by \Euclid was too early be able to detect the SN, but \Euclid observations do provide useful information on the host. In Fig.~\ref{fig:host} we compare the \Euclid $I_{\scriptscriptstyle\rm{E}}$ band observations with the ZTF detection image. As would be expected given the differences in their respective imaging specifications, \Euclid provides vastly enhanced detail. A more useful comparison is with the Legacy Survey DR10 image \citep{Dey2019} at the same location. Legacy survey images have similar resolution to that anticipated from Rubin (0\arcsecf262 and 0\arcsecf2 respectively) making them ideal for comparison and to highlight the benefit that \Euclid data would bring to LSST detected transients. Once again the improvement in detail between \Euclid and Legacy Survey can be readily observed with substantially more structure visible in the \Euclid image.

This improvement in detail goes beyond the aesthetic as it allows for improved morphology measurements, using tools such as \texttt{GALFIT} \citep{2010AJ....139.2097P}. This allows for some aspects of the higher level products proposed in \citet{2022zndo...5836022G} to be produced, specifically DDP-33 transient host parameters. Although this goes beyond the scope of this work it demonstrates the necessity and benefit of having an underlying automated matching system to bring data together. 

\subsubsection{ZTF24abxmfgs}
\begin{figure}
    \centering
    \includegraphics[width=\linewidth]{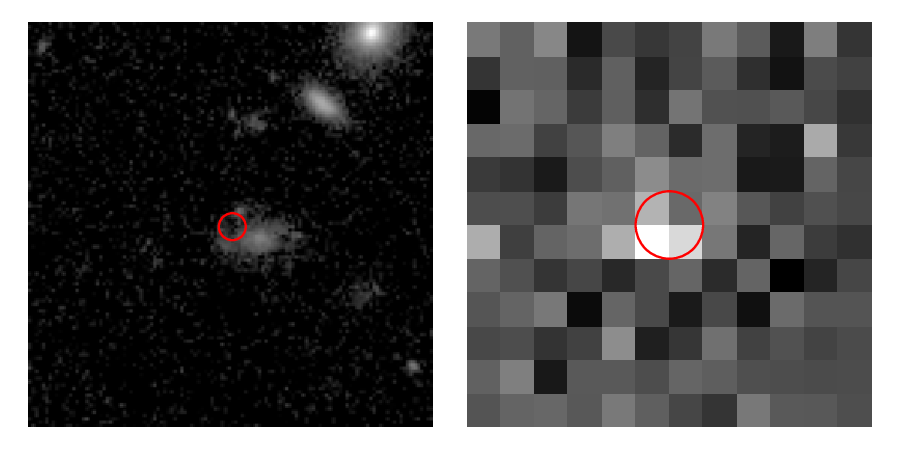}
    \caption{\textit{Left:} \Euclid $I_{\scriptscriptstyle\rm{E}}$ band $12\arcsec\times12\arcsec$ image at the location of ZTF24abxmfgs with potential host directly to the right of the highlighted coordinates. \textit{Right}: ZTF \textit{r}-band discovery image of ZTF24abxmfgs on the same scale; in both North on top and East to the left.}%
    \label{fig:orphan}
\end{figure}

ZTF24abxmfgs is transient source identified by Sherlock to be an ``orphan'' i.e., without a known host galaxy. It was discovered by ZTF on 2024-12-09, having previously been observed by \Euclid on 2024-10-20. Once again this observation was too early to make an observation of the transient, but \Euclid did identify a possible host. In Fig.~\ref{fig:orphan} we show both the ZTF discovery image for ZTF24abxmfgs and the Euclid $I_{\scriptscriptstyle\rm{E}}$ band image at the same location.

In the \Euclid image there appears to be a faint galaxy just to the west of the transient location, which on inspection would imply that the transient originated on the edge of this galaxy's disc. This allows not only for a host association to be added to sources such as this but to also associate host properties. Adding this information is useful when using Type Ia SNe for cosmology, as corrections to SN distance measurements can be derived from host properties \citep{2025A&A...694A..14S}.

Although Rubin will have similar photometric depth to \Euclid (especially after multiple epochs are stacked) which will likely reduce the frequency of Rubin ``orphan'' transients that \Euclid can identify a host for, there will still be cases where this occurs. This will be most prominent in the cases where there are particularly red or compact hosts.

\section{Next steps}\label{sec:nextSteps}
There are several improvements that we hope to take with this work in the future. First and foremost is completing the switch from using ZTF to using Rubin as the source of the transient alerts to match. This will not happen until the Rubin alert stream starts operations and Lasair starts serving alerts from it. This is expected to occur in early 2026. Only minor changes will be required to manage the transition, namely expanding the user interface to show 5 Rubin bands instead of 2 ZTF bands alongside some similar changes to back end functions. We expect that due to the standardisation of the Lasair alert packets between ZTF and Rubin operations we would be able to complete the transition quasi-simultaneously with the start of the Rubin alert stream.

The primary challenge of moving to Rubin from ZTF will be the increase in transients detected, and it will be key to ensure that the infrastructure of the system scales suitably under this increased load. As previously stated we expect 10 million alerts per night from Rubin, however many of these will be associated with previously detected transients e.g. a further observation of a supernova; a variable object e.g. a variable star, outbursting cataclysmic variable, or AGN; or Solar System objects. Assuming Rubin observes out to $z=1$, then from \citet[][figure 11]{2014AJ....148...13R} we have an intrinsic SN Ia rate of $\sim~4\times10^{-5}\, \rm year^{-1} \rm Mpc^{-3}$. For the $\sim18\,000\,\rm deg^{2}$ coverage of Rubin this corresponds to a volume of $6\times10^{10}\,\rm Mpc^{3}$ and thus a SN Ia rate of $\sim2.4\times10^{6}\,\rm year^{-1}$. As SN Ia make up approximately $30\%$ of the SN population \citep{Bulbul_2012}, and by extension the transient population (neglecting variable stars) due to the relatively small numbers of non-SN transients, we can estimate that the rate of unique transients observed by Rubin will be $\sim 8\times10^{6}\,\rm year^{-1}$. The overlap between the \Euclid survey area and Rubin is $\sim12\,000\,\rm deg^{2}$ which gives $\sim 5.3\times10^{6}$ transients per year originating from Rubin which could potentially appear in both surveys, or at very least the host could be studied.

By using the relative frequency of transient types and assuming the typical duration that a transient would be visible by \Euclid, we can estimate the number of different types of Rubin-detected transients that \Euclid is likely to observe: SN Ia, $30\%$ and $\sim50\,\rm d$ \citep{Bulbul_2012,2024arXiv241109740B}; other SNe (e.g. core collapse), $\sim70\%$ and $\sim100\,\rm d$ \citep{1997ARA&A..35..309F}; tidal disruption events (TDE), $\sim0.05\%$ and $\sim150\,\rm d$ \citep{2024A&A...690A.384S,2025A&A...697A.159G}; and kilonova, $\sim0.02\%$ and $\sim7\,\rm d$ \citep{2019MNRAS.485.4260S,2010MNRAS.406.2650M}. We further assume that Rubin transient detections are uniform over time and sky and that \Euclid observations are uncorrelated to Rubin observations (this is untrue for EDF-South, where there is planned coordination).

Forecasts of gravitationally lensed SNe (glSNe) in the Rubin footprint predict 44 glSN Ia per year \citep[][]{2024MNRAS.531.3509A}, and 88 glSNe per year, all types included \citep[][]{2024MNRAS.535.2523S}, from which 44 should provide images of the lensed host. Additional forecasts provided by \citet{2019MNRAS.487.3342W} and by \citet{2019ApJS..243....6G} give higher prediction rates depending on the method they used. We use the numbers provided by \citet[][]{2024MNRAS.531.3509A} and by \citet[][]{2024MNRAS.535.2523S} to estimate the rates of glSNe assuming the time delay will, on average, provide a longer visibility of 180$\%$ for both SN Ia and Other SN types. Finally, we point out that \Euclid is a lensing galaxy-galaxy systems discovery engine. It will continue to increase substantially the number of good lens candidates \citep[see e.g.][and the series of Q1 papers dedicated to lensing discoveries]{Q1-SP048} and will also provide to the discovery or follow-up of lensed transients, hence contributing to better time delay estimates of glSNe, which are necessary to get accurate estimates of \textit{H$_{0}$} \citep[for a recent review see][]{2024SSRv..220...13S}. 

In \autoref{tab:rates} we show the yearly number of transients originating from Rubin that \Euclid would be able to observe broken down by transient type for both the EWS and the EDFs. 

\begin{table}
    \centering
    \caption{Estimated number of Rubin originating transients observable by \Euclid in the wide survey and deep fields per year}
    \label{tab:rates}
    \begin{tabular}{ccc}
    \hline
    \hline
    Type & EWS & EDFs \\
    \hline
    SN Ia & $3.6\times10^4$ & $3.0\times10^3$\\
    Other SNe & $1.7\times10^5$ & $1.4\times10^4$ \\
    TDE & 180 & 7 \\ 
    Kilonova & 3 & 0.3 \\
    glSN Ia & $1.0 $ & $1.5$ \\
    glSNe Other & $4.0 $ & $6.0$ \\
    \hline
    \end{tabular}
\end{table}

We also look forward to integrating with a unified \Euclid transients database which would allow us to leverage the benefits that it would bring as discussed in Sect. \ref{sec:strategy}. This would require additional database look up steps to be implemented as well as a version of the \Euclid transient detection pipeline to be integrated with our code base. It is also possible that we would need to restructure the database which we operate on here to have a better unified structure. Whilst this has the potential to be a substantial project to complete, it would greatly enhance the effectiveness of this work and that of the \Euclid transient stream more broadly by lending better context to detections. Furthermore it would allow us to handle multi-epoch observations more accurately by using the difference imaging photometry that the transient pipeline will provide. 

The current form of this project has been to create a web app. This makes it very easy for a user to access the project and identify transients which have been matched and are worthy of further study. However it does not allow users to access any of the data, e.g., the photometry, outside the provided interface. Creating an API accessed through an associated python module would allow users to integrate aspects of this project directly into their workflow. For example being able to download the contents of the alert detail page or being able to run the user-defined matching (Sect. \ref{sec:userDefined}) directly from a script or command line would allow a user to search for and inspect objects arising from other aspects of their research.

Another potential future avenue is to create a Lasair annotator. This is a feature offered by Lasair where community users can add information back to an alert based on their own analysis. There is potential to use the output of work here to annotate Lasair alerts with information from \Euclid. This could range from a simple confirmation of an approximately contemporaneous \Euclid observation through to supplying photometric information. The exact nature that this takes will depend on future decisions relating to proprietary data rights for \Euclid in the transient science domain.

\subsection{Full joint transient DDPs}
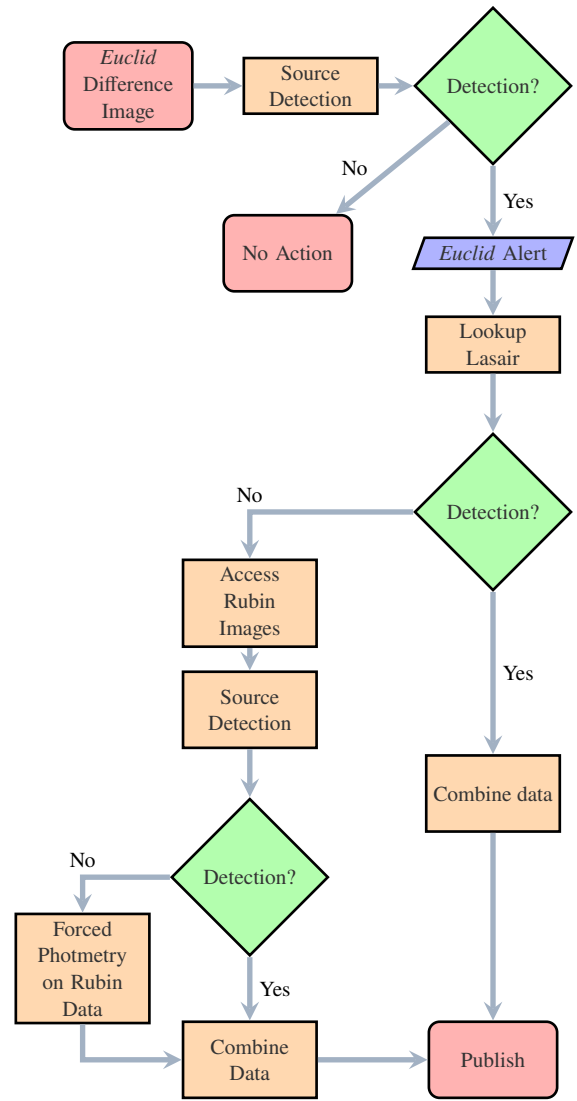
\begin{figure}
\centering
\begin{tikzpicture}[node distance=1.2cm]
\node (Start) [512bdd77-c3aa-5669-a956-85f7a90c6fb4] {\Euclid Difference Image};
\node (SExtractor) [69bbb168-da59-5865-902f-94e77902bf95, right of=Start, xshift=1.2cm] {Source Detection};
\node (detect) [b0e005f8-e267-5a3a-a638-881fb9faed1d, right of=SExtractor, xshift=1.2cm] {Detection?};
\node (88dd10f0-a3ee-4c37-8698-bb80a8d017dc) [7f27e395-8473-5b8a-b568-31425bce482a, below of=detect, yshift=-1.0cm] {\Euclid Alert};
\node (noDetect) [512bdd77-c3aa-5669-a956-85f7a90c6fb4, left of=88dd10f0-a3ee-4c37-8698-bb80a8d017dc, xshift=-1.5cm] {No Action};
\node (e448dfac-a5b8-484f-a6a6-d4f2ec29d861) [69bbb168-da59-5865-902f-94e77902bf95, below of=88dd10f0-a3ee-4c37-8698-bb80a8d017dc] {Lookup Lasair};
\node (6aa237f9-fe3b-404c-b3b7-2a948fdc0972) [b0e005f8-e267-5a3a-a638-881fb9faed1d, below of=e448dfac-a5b8-484f-a6a6-d4f2ec29d861, yshift=-1.0cm] {Detection?};
\node (f9624a31-43e7-4bb9-a699-54faa72c5a2d) [b4eeb85a-e852-5337-aaeb-b1b7ed281176, left of=6aa237f9-fe3b-404c-b3b7-2a948fdc0972, xshift=-2.0cm, yshift=-1.2cm] {Access Rubin Images};
\node (da355d7e-1ee7-4027-b7dd-677782c9f033) [b4eeb85a-e852-5337-aaeb-b1b7ed281176, below of=f9624a31-43e7-4bb9-a699-54faa72c5a2d, xshift=0cm, yshift=-0.2cm] {Source Detection};
\node (detectRubin) [b0e005f8-e267-5a3a-a638-881fb9faed1d, below of=da355d7e-1ee7-4027-b7dd-677782c9f033, yshift=-1.0cm] {Detection?};
\node (nonDetect) [b4eeb85a-e852-5337-aaeb-b1b7ed281176, left of=detectRubin, yshift=-1.2cm, xshift=-1.0cm] {Forced Photmetry on Rubin Data};

\node (8f3b6556-3a67-42f6-bb70-0c50d3250361) [b4eeb85a-e852-5337-aaeb-b1b7ed281176, below of=detectRubin, yshift=-1.2cm] {Combine Data};
\node (809d5e60-ab79-4849-bc8a-241fba129076) [574b93ce-5d31-5729-9ed0-9a79e73d4725, below of=6aa237f9-fe3b-404c-b3b7-2a948fdc0972, yshift=-2.5cm] {Combine data};
\node (5b0051da-629f-4ef8-a91c-3bb0340ff2b5) [512bdd77-c3aa-5669-a956-85f7a90c6fb4, below of=6aa237f9-fe3b-404c-b3b7-2a948fdc0972, yshift=-6.0cm] {Publish};

\draw [7be24b85-97d0-5b76-ba9e-d94005dca8f2] (88dd10f0-a3ee-4c37-8698-bb80a8d017dc) --  (e448dfac-a5b8-484f-a6a6-d4f2ec29d861);
\draw [7be24b85-97d0-5b76-ba9e-d94005dca8f2] (6aa237f9-fe3b-404c-b3b7-2a948fdc0972) --  node[anchor=west] {Yes} (809d5e60-ab79-4849-bc8a-241fba129076);
\draw [7be24b85-97d0-5b76-ba9e-d94005dca8f2] (e448dfac-a5b8-484f-a6a6-d4f2ec29d861) --  (6aa237f9-fe3b-404c-b3b7-2a948fdc0972);
\draw [7be24b85-97d0-5b76-ba9e-d94005dca8f2] (6aa237f9-fe3b-404c-b3b7-2a948fdc0972) -|  node[anchor=south] {No} (f9624a31-43e7-4bb9-a699-54faa72c5a2d);
\draw [7be24b85-97d0-5b76-ba9e-d94005dca8f2] (809d5e60-ab79-4849-bc8a-241fba129076) --  (5b0051da-629f-4ef8-a91c-3bb0340ff2b5);
\draw [7be24b85-97d0-5b76-ba9e-d94005dca8f2] (f9624a31-43e7-4bb9-a699-54faa72c5a2d) --  (da355d7e-1ee7-4027-b7dd-677782c9f033);
\draw [7be24b85-97d0-5b76-ba9e-d94005dca8f2] (da355d7e-1ee7-4027-b7dd-677782c9f033) --  (detectRubin);
\draw [7be24b85-97d0-5b76-ba9e-d94005dca8f2] (8f3b6556-3a67-42f6-bb70-0c50d3250361) --  (5b0051da-629f-4ef8-a91c-3bb0340ff2b5);
\draw [7be24b85-97d0-5b76-ba9e-d94005dca8f2] (detectRubin) --  node[anchor=west] {Yes} (8f3b6556-3a67-42f6-bb70-0c50d3250361);
\draw [7be24b85-97d0-5b76-ba9e-d94005dca8f2] (detectRubin) -| node[anchor=south] {No} (nonDetect);
\draw [7be24b85-97d0-5b76-ba9e-d94005dca8f2] (detect) -- node[anchor=east, xshift=-0.2cm] {No} (noDetect);
\draw [7be24b85-97d0-5b76-ba9e-d94005dca8f2] (detect) --  node[anchor=west] {Yes} (88dd10f0-a3ee-4c37-8698-bb80a8d017dc);
\draw [7be24b85-97d0-5b76-ba9e-d94005dca8f2] (SExtractor) --  (detect);
\draw [7be24b85-97d0-5b76-ba9e-d94005dca8f2] (Start) --  (SExtractor);
\draw [7be24b85-97d0-5b76-ba9e-d94005dca8f2] (nonDetect) |-  (8f3b6556-3a67-42f6-bb70-0c50d3250361);
\end{tikzpicture}
\caption{Schematic flow of operations in creating DDP following a \Euclid Alert}\label{fig:strat2}
\end{figure}

The ultimate goal of this work is to serve as one of two necessary aspects to create joint derived data products for transient science using \Euclid and Rubin data. Whilst we have focused here on the case of a transient alert originating from Rubin data (which will be by far the more common case), the other aspect is the case where a transient is first detected by \Euclid. In Fig.~\ref{fig:strat2} we show a schematic of how starting at a \Euclid difference image a similar joint data product to those presented here might be generated. When compared with Fig.~\ref{fig:flowchart} it should be immediately apparent that much of the logic required is similar, thus when considering this aspect of the DDP the work that we present here could be viewed as a path finder.

Although there will be a lot of surface level commonality between the two sides of the product, development of the products arising from \Euclid alerts will require a substantial effort to realise as there are substantial differences in the Rubin data that need to be considered, namely the provided data products and the volume of data products that need to be handled. For example, it may not be tractable to run the automatic matching on the same temporal overlap that we use in this work. Equally, with far more repeat visits in a shorter period of time in Rubin data, it will be necessary to consider how best to present this data, for example the ability to blink between science and difference images is a feature that would likely be desired. In addition to building a product for this case it will also be necessary to remain mindful of how both sides will integrate together and what aspects of re-engineering will be required to this work so that both aspects of the DDP process can work seamlessly and consistently together.

\subsection{Beyond Transient DDPs}

In addition to the transient DDP products there are several other time domain products proposed in \citet{2022zndo...5836022G} which require the rapid production of joint light curves and/or cutouts, where the system presented here could be the basis for. For example, DDP 47 calls for prompt light curves of extreme AGN variability events, such as TDEs. Many of these events will be reported as Rubin alerts and passed to Lasair, which is already capable of identifying the origin of these events; thus making it straightforward to integrate them in this system. As highlighted by \citet{2022zndo...5836022G} when working with AGN there will be the particular challenge of deblending the AGN and host galaxy light. Some tools already exist to be able to do this e.g. \texttt{SCARLET} and \texttt{SCARLET 2} \citep{2018A&C....24..129M,2024A&C....4900875S} and these potentially could be integrated to achieve this. This may have the added benefit of helping reduce the impact of host galaxy light contamination in the primary transient science products discussed, especially where we have highlighted the impact of host galaxy contamination.

Other time domain products which could be produced using this system as a starting point are DDP 7 and 8, Solar System object cutouts and lightcurves. This would be a more involved evolution of the system as Lasair does not support Solar System objects, and would require integrating another LSST community broker such as the Solar System Notification Alert Processing System \citep[SNAPS][]{2023AJ....165..111T} which specialises in Solar System objects.

\section{Conclusions}
We have presented a prototype system for combining Rubin transient alerts with \Euclid data, using ZTF data alerts provided by the LSST:UK transient alert broker Lasair as a pathfinder. We have demonstrated both the technical capacity of the system to process and match incoming alerts before producing useable science products. Beyond this we have highlighted a number of potential science use cases of the system including early time observations of transients prior to ground-based detection, and host galaxy detection and study. With the start of Rubin science operations in 2026 we hope to make the switch from ZTF data to Rubin and make this tool available to the DDP transient community.

\section*{Acknowledgements}

\AckECon\\
\AckQone\\
This work was funded by UKRI grant ST/Y005511/1. For the purpose of open access, the authors have applied a creative commons attribution (CC BY) licence to any author accepted manuscript version arising.\\
VP acknowledge support from MINIGRANT Euclid Transient Search RSN2 1.05.23.04.02\\
CGC and FP acknowledge support from the Spanish Ministerio de Ciencia, Innovación y Universidades (MICINN) under grant numbers PID2022-141915NB-C21.\\
This work made use of Astropy:\footnote{\href{http://www.astropy.org}{http://www.astropy.org}} a community-developed core Python package and an ecosystem of tools and resources for astronomy \citep{astropy:2013, astropy:2018, astropy:2022}.
This research made use of Photutils, an Astropy package for detection and photometry of astronomical sources \citep{larry_bradley_2025_14606896}.\\
The Legacy Surveys consist of three individual and complementary projects: the Dark Energy Camera Legacy Survey (DECaLS; Proposal ID \#~2014B-0404; PIs: David Schlegel and Arjun Dey), the Beijing-Arizona Sky Survey (BASS; NOAO Prop. ID \#~2015A-0801; PIs: Zhou Xu and Xiaohui Fan), and the Mayall z-band Legacy Survey (MzLS; Prop. ID \#~2016A-0453; PI: Arjun Dey). DECaLS, BASS and MzLS together include data obtained, respectively, at the Blanco telescope, Cerro Tololo Inter-American Observatory, NSF’s NOIRLab; the Bok telescope, Steward Observatory, University of Arizona; and the Mayall telescope, Kitt Peak National Observatory, NOIRLab. Pipeline processing and analyses of the data were supported by NOIRLab and the Lawrence Berkeley National Laboratory (LBNL). The Legacy Surveys project is honored to be permitted to conduct astronomical research on Iolkam Du’ag (Kitt Peak), a mountain with particular significance to the Tohono O’odham Nation.\\
NOIRLab is operated by the Association of Universities for Research in Astronomy (AURA) under a cooperative agreement with the National Science Foundation. LBNL is managed by the Regents of the University of California under contract to the U.S. Department of Energy.\\
This project used data obtained with the Dark Energy Camera (DECam), which was constructed by the Dark Energy Survey (DES) collaboration. Funding for the DES Projects has been provided by the U.S. Department of Energy, the U.S. National Science Foundation, the Ministry of Science and Education of Spain, the Science and Technology Facilities Council of the United Kingdom, the Higher Education Funding Council for England, the National Center for Supercomputing Applications at the University of Illinois at Urbana-Champaign, the Kavli Institute of Cosmological Physics at the University of Chicago, Center for Cosmology and Astro-Particle Physics at the Ohio State University, the Mitchell Institute for Fundamental Physics and Astronomy at Texas A\& M University, Financiadora de Estudos e Projetos, Fundacao Carlos Chagas Filho de Amparo, Financiadora de Estudos e Projetos, Fundacao Carlos Chagas Filho de Amparo a Pesquisa do Estado do Rio de Janeiro, Conselho Nacional de Desenvolvimento Cientifico e Tecnologico and the Ministerio da Ciencia, Tecnologia e Inovacao, the Deutsche Forschungsgemeinschaft and the Collaborating Institutions in the Dark Energy Survey. The Collaborating Institutions are Argonne National Laboratory, the University of California at Santa Cruz, the University of Cambridge, Centro de Investigaciones Energeticas, Medioambientales y Tecnologicas-Madrid, the University of Chicago, University College London, the DES-Brazil Consortium, the University of Edinburgh, the Eidgenossische Technische Hochschule (ETH) Zurich, Fermi National Accelerator Laboratory, the University of Illinois at Urbana-Champaign, the Institut de Ciencies de l’Espai (IEEC/CSIC), the Institut de Fisica d’Altes Energies, Lawrence Berkeley National Laboratory, the Ludwig Maximilians Universitat Munchen and the associated Excellence Cluster Universe, the University of Michigan, NSF’s NOIRLab, the University of Nottingham, the Ohio State University, the University of Pennsylvania, the University of Portsmouth, SLAC National Accelerator Laboratory, Stanford University, the University of Sussex, and Texas A\&M University.\\
BASS is a key project of the Telescope Access Program (TAP), which has been funded by the National Astronomical Observatories of China, the Chinese Academy of Sciences (the Strategic Priority Research Program “The Emergence of Cosmological Structures” Grant \# XDB09000000), and the Special Fund for Astronomy from the Ministry of Finance. The BASS is also supported by the External Cooperation Program of Chinese Academy of Sciences (Grant \# ~114A11KYSB20160057), and Chinese National Natural Science Foundation (Grant \#~12120101003, \#~11433005).\\
The Legacy Survey team makes use of data products from the Near-Earth Object Wide-field Infrared Survey Explorer (NEOWISE), which is a project of the Jet Propulsion Laboratory/California Institute of Technology. NEOWISE is funded by the National Aeronautics and Space Administration.\\
The Legacy Surveys imaging of the DESI footprint is supported by the Director, Office of Science, Office of High Energy Physics of the U.S. Department of Energy under Contract No. DE-AC02-05CH1123, by the National Energy Research Scientific Computing Center, a DOE Office of Science User Facility under the same contract; and by the U.S. National Science Foundation, Division of Astronomical Sciences under Contract No. AST-0950945 to NOAO.\\
Lasair is supported by the UKRI Science and Technology Facilities Council and is a collaboration between the University of Edinburgh (grant ST/N002512/1) and Queen’s University Belfast (grant ST/N002520/1) within the LSST:UK Science Consortium. ZTF is supported by National Science Foundation grant AST-1440341 and a collaboration including Caltech, IPAC, the Weizmann Institute for Science, the Oskar Klein Center at Stockholm University, the University of Maryland, the University of Washington, Deutsches Elektronen-Synchrotron and Humboldt University, Los Alamos National Laboratories, the TANGO Consortium of Taiwan, the University of Wisconsin at Milwaukee, and Lawrence Berkeley National Laboratories. Operations are conducted by COO, IPAC, and UW. This research has made use of ``Aladin sky atlas'' developed at CDS, Strasbourg Observatory, France 2000A\&AS..143…33B and 2014ASPC..485..277B.\\
We thank the referees for the time that they have taken to provide comments and advice on this work.

\section*{Data Availability}

\Euclid data is available as part of the Euclid Q1 Data release; \href{https://www.cosmos.esa.int/web/euclid/q1-data}{cosmos.esa.int/web/euclid/q1-data}\\
ZTF photometry and cutouts are available from Lasair; \href{http://lasair-ztf.lsst.ac.uk}{lasair-ztf.lsst.ac.uk}\\
Legacy Survey Cutouts are available at: \href{https://www.legacysurvey.org/viewer/urls}{legacysurvey.org/viewer/urls}

\section*{Conflict of Interest Statement}

Authors declare no conflict of interest.

\section*{Affiliations}
$^{1}$Department of Physics, Lancaster University, Lancaster, LA1 4YB, UK\\
$^{2}$Instituto de Astrof\'{\i}sica de Canarias, E-38205 La Laguna, Tenerife, Spain\\
$^{3}$Universidad de La Laguna, Dpto. Astrof\'\i sica, E-38206 La Laguna, Tenerife, Spain\\
$^{4}$Max-Planck-Institut f\"ur Astronomie, K\"onigstuhl 17, 69117 Heidelberg, Germany\\
$^{5}$INAF-Osservatorio Astronomico di Capodimonte, Via Moiariello 16, 80131 Napoli, Italy\\
$^{6}$Dipartimento di Fisica "E. Pancini", Universita degli Studi di Napoli Federico II, Via Cinthia 6, 80126, Napoli, Italy\\
$^{7}$National Astronomical Observatory of Japan, National Institutes of Natural Sciences, 2-21-1 Osawa, Mitaka, Tokyo 181-8588, Japan\\
$^{8}$Graduate Institute for Advanced Studies, SOKENDAI, 2-21-1 Osawa, Mitaka, Tokyo 181-8588, Japan\\
$^{9}$School of Physics and Astronomy, Monash University, Clayton, VIC 3800, Australia\\
$^{10}$Department of Physics and Astronomy, Vesilinnantie 5, University of Turku, 20014 Turku, Finland\\
$^{11}$ESAC/ESA, Camino Bajo del Castillo, s/n., Urb. Villafranca del Castillo, 28692 Villanueva de la Ca\~nada, Madrid, Spain\\
$^{12}$School of Mathematics and Physics, University of Surrey, Guildford, Surrey, GU2 7XH, UK\\
$^{13}$INAF-Osservatorio Astronomico di Brera, Via Brera 28, 20122 Milano, Italy\\
$^{14}$INAF-Osservatorio di Astrofisica e Scienza dello Spazio di Bologna, Via Piero Gobetti 93/3, 40129 Bologna, Italy\\
$^{15}$IFPU, Institute for Fundamental Physics of the Universe, via Beirut 2, 34151 Trieste, Italy\\
$^{16}$INAF-Osservatorio Astronomico di Trieste, Via G. B. Tiepolo 11, 34143 Trieste, Italy\\
$^{17}$INFN, Sezione di Trieste, Via Valerio 2, 34127 Trieste TS, Italy\\
$^{18}$SISSA, International School for Advanced Studies, Via Bonomea 265, 34136 Trieste TS, Italy\\
$^{19}$Dipartimento di Fisica e Astronomia, Universit\`a di Bologna, Via Gobetti 93/2, 40129 Bologna, Italy\\
$^{20}$INFN-Sezione di Bologna, Viale Berti Pichat 6/2, 40127 Bologna, Italy\\
$^{21}$INAF-Osservatorio Astronomico di Padova, Via dell'Osservatorio 5, 35122 Padova, Italy\\
$^{22}$Dipartimento di Fisica, Universit\`a di Genova, Via Dodecaneso 33, 16146, Genova, Italy\\
$^{23}$INFN-Sezione di Genova, Via Dodecaneso 33, 16146, Genova, Italy\\
$^{24}$Department of Physics "E. Pancini", University Federico II, Via Cinthia 6, 80126, Napoli, Italy\\
$^{25}$Dipartimento di Fisica, Universit\`a degli Studi di Torino, Via P. Giuria 1, 10125 Torino, Italy\\
$^{26}$INFN-Sezione di Torino, Via P. Giuria 1, 10125 Torino, Italy\\
$^{27}$INAF-Osservatorio Astrofisico di Torino, Via Osservatorio 20, 10025 Pino Torinese (TO), Italy\\
$^{28}$Institute for Astronomy, University of Edinburgh, Royal Observatory, Blackford Hill, Edinburgh EH9 3HJ, UK\\
$^{29}$Leiden Observatory, Leiden University, Einsteinweg 55, 2333 CC Leiden, The Netherlands\\
$^{30}$INAF-IASF Milano, Via Alfonso Corti 12, 20133 Milano, Italy\\
$^{31}$Centro de Investigaciones Energ\'eticas, Medioambientales y Tecnol\'ogicas (CIEMAT), Avenida Complutense 40, 28040 Madrid, Spain\\
$^{32}$Port d'Informaci\'{o} Cient\'{i}fica, Campus UAB, C. Albareda s/n, 08193 Bellaterra (Barcelona), Spain\\
$^{33}$Institute for Theoretical Particle Physics and Cosmology (TTK), RWTH Aachen University, 52056 Aachen, Germany\\
$^{34}$Deutsches Zentrum f\"ur Luft- und Raumfahrt e. V. (DLR), Linder H\"ohe, 51147 K\"oln, Germany\\
$^{35}$INAF-Osservatorio Astronomico di Roma, Via Frascati 33, 00078 Monteporzio Catone, Italy\\
$^{36}$INFN section of Naples, Via Cinthia 6, 80126, Napoli, Italy\\
$^{37}$Institute for Astronomy, University of Hawaii, 2680 Woodlawn Drive, Honolulu, HI 96822, USA\\
$^{38}$Dipartimento di Fisica e Astronomia "Augusto Righi" - Alma Mater Studiorum Universit\`a di Bologna, Viale Berti Pichat 6/2, 40127 Bologna, Italy\\
$^{39}$Jodrell Bank Centre for Astrophysics, Department of Physics and Astronomy, University of Manchester, Oxford Road, Manchester M13 9PL, UK\\
$^{40}$European Space Agency/ESRIN, Largo Galileo Galilei 1, 00044 Frascati, Roma, Italy\\
$^{41}$Universit\'e Claude Bernard Lyon 1, CNRS/IN2P3, IP2I Lyon, UMR 5822, Villeurbanne, F-69100, France\\
$^{42}$Institut de Ci\`{e}ncies del Cosmos (ICCUB), Universitat de Barcelona (IEEC-UB), Mart\'{i} i Franqu\`{e}s 1, 08028 Barcelona, Spain\\
$^{43}$Instituci\'o Catalana de Recerca i Estudis Avan\c{c}ats (ICREA), Passeig de Llu\'{\i}s Companys 23, 08010 Barcelona, Spain\\
$^{44}$Institut de Ciencies de l'Espai (IEEC-CSIC), Campus UAB, Carrer de Can Magrans, s/n Cerdanyola del Vall\'es, 08193 Barcelona, Spain\\
$^{45}$UCB Lyon 1, CNRS/IN2P3, IUF, IP2I Lyon, 4 rue Enrico Fermi, 69622 Villeurbanne, France\\
$^{46}$Mullard Space Science Laboratory, University College London, Holmbury St Mary, Dorking, Surrey RH5 6NT, UK\\
$^{47}$Universit\'e Paris-Saclay, Universit\'e Paris Cit\'e, CEA, CNRS, AIM, 91191, Gif-sur-Yvette, France\\
$^{48}$Department of Astronomy, University of Geneva, ch. d'Ecogia 16, 1290 Versoix, Switzerland\\
$^{49}$Universit\'e Paris-Saclay, CNRS, Institut d'astrophysique spatiale, 91405, Orsay, France\\
$^{50}$INFN-Padova, Via Marzolo 8, 35131 Padova, Italy\\
$^{51}$Aix-Marseille Universit\'e, CNRS/IN2P3, CPPM, Marseille, France\\
$^{52}$INAF-Istituto di Astrofisica e Planetologia Spaziali, via del Fosso del Cavaliere, 100, 00100 Roma, Italy\\
$^{53}$University Observatory, LMU Faculty of Physics, Scheinerstr.~1, 81679 Munich, Germany\\
$^{54}$Max Planck Institute for Extraterrestrial Physics, Giessenbachstr. 1, 85748 Garching, Germany\\
$^{55}$Universit\"ats-Sternwarte M\"unchen, Fakult\"at f\"ur Physik, Ludwig-Maximilians-Universit\"at M\"unchen, Scheinerstr.~1, 81679 M\"unchen, Germany\\
$^{56}$Institute of Theoretical Astrophysics, University of Oslo, P.O. Box 1029 Blindern, 0315 Oslo, Norway\\
$^{57}$Jet Propulsion Laboratory, California Institute of Technology, 4800 Oak Grove Drive, Pasadena, CA, 91109, USA\\
$^{58}$Felix Hormuth Engineering, Goethestr. 17, 69181 Leimen, Germany\\
$^{59}$Technical University of Denmark, Elektrovej 327, 2800 Kgs. Lyngby, Denmark\\
$^{60}$Cosmic Dawn Center (DAWN), Denmark\\
$^{61}$NASA Goddard Space Flight Center, Greenbelt, MD 20771, USA\\
$^{62}$Universit\'e de Gen\`eve, D\'epartement de Physique Th\'eorique and Centre for Astroparticle Physics, 24 quai Ernest-Ansermet, CH-1211 Gen\`eve 4, Switzerland\\
$^{63}$Department of Physics, P.O. Box 64, University of Helsinki, 00014 Helsinki, Finland\\
$^{64}$Helsinki Institute of Physics, Gustaf H{\"a}llstr{\"o}min katu 2, University of Helsinki, 00014 Helsinki, Finland\\
$^{65}$Laboratoire d'etude de l'Univers et des phenomenes eXtremes, Observatoire de Paris, Universit\'e PSL, Sorbonne Universit\'e, CNRS, 92190 Meudon, France\\
$^{66}$SKAO, Jodrell Bank, Lower Withington, Macclesfield SK11 9FT, UK\\
$^{67}$Centre de Calcul de l'IN2P3/CNRS, 21 avenue Pierre de Coubertin 69627 Villeurbanne Cedex, France\\
$^{68}$Dipartimento di Fisica "Aldo Pontremoli", Universit\`a degli Studi di Milano, Via Celoria 16, 20133 Milano, Italy\\
$^{69}$INFN-Sezione di Milano, Via Celoria 16, 20133 Milano, Italy\\
$^{70}$University of Applied Sciences and Arts of Northwestern Switzerland, School of Computer Science, 5210 Windisch, Switzerland\\
$^{71}$Universit\"at Bonn, Argelander-Institut f\"ur Astronomie, Auf dem H\"ugel 71, 53121 Bonn, Germany\\
$^{72}$INFN-Sezione di Roma, Piazzale Aldo Moro, 2 - c/o Dipartimento di Fisica, Edificio G. Marconi, 00185 Roma, Italy\\
$^{73}$Aix-Marseille Universit\'e, CNRS, CNES, LAM, Marseille, France\\
$^{74}$Dipartimento di Fisica e Astronomia "Augusto Righi" - Alma Mater Studiorum Universit\`a di Bologna, via Piero Gobetti 93/2, 40129 Bologna, Italy\\
$^{75}$Department of Physics, Institute for Computational Cosmology, Durham University, South Road, Durham, DH1 3LE, UK\\
$^{76}$Universit\'e Paris Cit\'e, CNRS, Astroparticule et Cosmologie, 75013 Paris, France\\
$^{77}$CNRS-UCB International Research Laboratory, Centre Pierre Bin\'etruy, IRL2007, CPB-IN2P3, Berkeley, USA\\
$^{78}$Institut d'Astrophysique de Paris, 98bis Boulevard Arago, 75014, Paris, France\\
$^{79}$Institut d'Astrophysique de Paris, UMR 7095, CNRS, and Sorbonne Universit\'e, 98 bis boulevard Arago, 75014 Paris, France\\
$^{80}$Institute of Physics, Laboratory of Astrophysics, Ecole Polytechnique F\'ed\'erale de Lausanne (EPFL), Observatoire de Sauverny, 1290 Versoix, Switzerland\\
$^{81}$Telespazio UK S.L. for European Space Agency (ESA), Camino bajo del Castillo, s/n, Urbanizacion Villafranca del Castillo, Villanueva de la Ca\~nada, 28692 Madrid, Spain\\
$^{82}$Institut de F\'{i}sica d'Altes Energies (IFAE), The Barcelona Institute of Science and Technology, Campus UAB, 08193 Bellaterra (Barcelona), Spain\\
$^{83}$European Space Agency/ESTEC, Keplerlaan 1, 2201 AZ Noordwijk, The Netherlands\\
$^{84}$DARK, Niels Bohr Institute, University of Copenhagen, Jagtvej 155, 2200 Copenhagen, Denmark\\
$^{85}$Space Science Data Center, Italian Space Agency, via del Politecnico snc, 00133 Roma, Italy\\
$^{86}$Centre National d'Etudes Spatiales -- Centre spatial de Toulouse, 18 avenue Edouard Belin, 31401 Toulouse Cedex 9, France\\
$^{87}$Institute of Space Science, Str. Atomistilor, nr. 409 M\u{a}gurele, Ilfov, 077125, Romania\\
$^{88}$Dipartimento di Fisica e Astronomia "G. Galilei", Universit\`a di Padova, Via Marzolo 8, 35131 Padova, Italy\\
$^{89}$Institut f\"ur Theoretische Physik, University of Heidelberg, Philosophenweg 16, 69120 Heidelberg, Germany\\
$^{90}$Institut de Recherche en Astrophysique et Plan\'etologie (IRAP), Universit\'e de Toulouse, CNRS, UPS, CNES, 14 Av. Edouard Belin, 31400 Toulouse, France\\
$^{91}$Universit\'e St Joseph; Faculty of Sciences, Beirut, Lebanon\\
$^{92}$Departamento de F\'isica, FCFM, Universidad de Chile, Blanco Encalada 2008, Santiago, Chile\\
$^{93}$Institut d'Estudis Espacials de Catalunya (IEEC),  Edifici RDIT, Campus UPC, 08860 Castelldefels, Barcelona, Spain\\
$^{94}$Satlantis, University Science Park, Sede Bld 48940, Leioa-Bilbao, Spain\\
$^{95}$Institute of Space Sciences (ICE, CSIC), Campus UAB, Carrer de Can Magrans, s/n, 08193 Barcelona, Spain\\
$^{96}$Department of Physics and Helsinki Institute of Physics, Gustaf H\"allstr\"omin katu 2, University of Helsinki, 00014 Helsinki, Finland\\
$^{97}$Centre for Electronic Imaging, Open University, Walton Hall, Milton Keynes, MK7~6AA, UK\\
$^{98}$Departamento de F\'isica, Faculdade de Ci\^encias, Universidade de Lisboa, Edif\'icio C8, Campo Grande, PT1749-016 Lisboa, Portugal\\
$^{99}$Instituto de Astrof\'isica e Ci\^encias do Espa\c{c}o, Faculdade de Ci\^encias, Universidade de Lisboa, Tapada da Ajuda, 1349-018 Lisboa, Portugal\\
$^{100}$Cosmic Dawn Center (DAWN)\\
$^{101}$Niels Bohr Institute, University of Copenhagen, Jagtvej 128, 2200 Copenhagen, Denmark\\
$^{102}$Universidad Polit\'ecnica de Cartagena, Departamento de Electr\'onica y Tecnolog\'ia de Computadoras,  Plaza del Hospital 1, 30202 Cartagena, Spain\\
$^{103}$INFN-Bologna, Via Irnerio 46, 40126 Bologna, Italy\\
$^{104}$Caltech/IPAC, 1200 E. California Blvd., Pasadena, CA 91125, USA\\
$^{105}$Instituto de F\'isica Te\'orica UAM-CSIC, Campus de Cantoblanco, 28049 Madrid, Spain\\
$^{106}$Aurora Technology for European Space Agency (ESA), Camino bajo del Castillo, s/n, Urbanizacion Villafranca del Castillo, Villanueva de la Ca\~nada, 28692 Madrid, Spain\\
$^{107}$Department of Mathematics and Physics E. De Giorgi, University of Salento, Via per Arnesano, CP-I93, 73100, Lecce, Italy\\
$^{108}$INFN, Sezione di Lecce, Via per Arnesano, CP-193, 73100, Lecce, Italy\\
$^{109}$INAF-Sezione di Lecce, c/o Dipartimento Matematica e Fisica, Via per Arnesano, 73100, Lecce, Italy\\
$^{110}$ICL, Junia, Universit\'e Catholique de Lille, LITL, 59000 Lille, France



\bibliographystyle{mnras}
\makeatletter
\def\@bibpagesprefix#1{}
\makeatother
\bibliography{references} 


\appendix


\bsp	
\label{lastpage}
\end{document}